\documentclass[a4paper,twocolumn,11pt,unpublished]{quantumarticle}
\pdfoutput=1
\usepackage[utf8]{inputenc}
\usepackage[english]{babel}
\usepackage[T1]{fontenc}
\usepackage{amsmath}
\usepackage{hyperref}

\usepackage{graphicx}
\usepackage{subfigure}
\usepackage{epstopdf}
\usepackage{graphicx}

\usepackage{amsfonts}
\usepackage{amssymb, color, bm}

\usepackage{currvita}

\usepackage{algorithmicx}
\usepackage[noend]{algpseudocode}
\usepackage{algorithm,setspace}

\usepackage{physics}

\usepackage{multicol}
\usepackage{lipsum}
\usepackage{mwe}

\DeclareMathOperator{\E}{\mathbb{E}}

\usepackage{tikz}
\usepackage{lipsum}

\begin{document}

\title{Time-Warping Invariant Quantum Recurrent Neural Networks via Quantum-Classical Adaptive Gating}

\author{Ivana Nikoloska}
\affiliation{Eindhoven University of Technology, Eindhoven, 5612 AP, The Netherlands}
\email{i.nikoloska@tue.nl}
\author{Osvaldo Simeone}
\affiliation{King's College London, Strand, London, WC2R 2LS, United Kingdom}
\author{Leonardo Banchi$^{\ast}$}
\affiliation{Department of Physics and Astronomy, University of Florence, via Sansone 1, 
I-50019, Sesto Fiorentino (FI), Italy}
\affiliation{INFN, Sezione di Firenze, via Sansone 1, I-50019, Sesto Fiorentino (FI), Italy}
\author{Petar Veli\v{c}kovi\'c$^{\ast}$}
\affiliation{DeepMind, London, United Kingdom}
\def\thefootnote{${\ast}$}\footnotetext{alphabetical order}

\maketitle

\begin{abstract}
Adaptive gating plays a key role in temporal data processing via classical recurrent neural networks (RNN), as it facilitates retention of  past information necessary to predict the future, providing a mechanism that preserves invariance to time warping transformations. 
This paper builds on quantum recurrent neural networks (QRNNs), a dynamic model with quantum memory, to introduce a novel class of temporal data processing  quantum models that preserve invariance to time-warping transformations of the (classical) input-output sequences. The model, referred to as \emph{time warping-invariant QRNN (TWI-QRNN)}, augments a QRNN with a quantum-classical adaptive gating mechanism that chooses whether to apply a parameterized unitary transformation at each time step as a function of the past samples of the input sequence via a classical recurrent model. The TWI-QRNN model class is derived  from first principles, and its capacity to successfully implement time-warping transformations is experimentally demonstrated on examples with classical or quantum dynamics.   
\end{abstract}

\section{Introduction}
 
Sequential data is prevalent in both classical and quantum systems. Indeed, tasks like natural language processing \cite{chowdhary2020natural, hirschberg2015advances} or simulating the dynamics of quantum systems \cite{georgescu2014quantum, mitarai2018quantum} require computing architectures that are capable of capturing temporal dependencies by retaining only the information about the past that is necessary to predict the future. This calls for \emph{adaptive gating} mechanisms that can forget in a data-driven manner, so as to selectively memorize and overwrite information  \cite{graves2012long, graves2013speech}.

{\color{black}{For classical processes}}, it was recently formally proved that gating mechanisms ensure \emph{invariance to time warping} \cite{tallec2018can, bronstein2021geometric}. Consider, for example, monitoring vitals of patients in intensive care, or the evolution of a quantum system. Time-series data that result from these processes are not naturally discrete -- they are in fact obtained by sampling a continuous signal. However, in practice, the sampling rate cannot always be controlled, and moreover, it may not always be fixed. Time warping transformations capture such dynamically changing sampling rates, or basic operations such as inserting a, possibly varying, number of zeros or white spaces between the elements of an input sequence. As such, time warping can model important practical aspects such as imperfect sampling due to jitter or clock drifts, or changes in time scales.  Thereby, it is imperative to develop models that are immune to these transformations and are able to fit the corresponding ``time-warped'' signals.

Conventional recurrent neural networks (RNNs), which do not implement gating mechanisms,  are not robust to time-warping transformations, failing even in simple settings with the additional of few zeros between samples \cite{tallec2018can}. In contrast, classical RNN models with gating, such as long short-term memories (LSTMs) and gated recurrent units (GRUs), can successfully adapt to transformed inputs. This paper aims at investigating quantum models that preserve time warping-invariance for  temporal data processing, targeting both classical and quantum dynamics. 

\subsection{Related Work}

Reflecting the underlying symmetries of a given data set in the choice of the inductive bias is well known to be of fundamental importance for the success of classical machine learning. Whilst symmetries have always played an imperative role in studying physical systems and understanding the laws of physics, the study of symmetries in data has just recently gained momentum, leading to the blossoming field of geometric deep learning \cite{bronstein2021geometric}. Symmetries  formalize the invariance of objects under some set of operations. For example, the binding energy of a molecule does not change by permuting the order of the atoms, and a picture of a cat still depicts a cat regardless of the position of the cat within the image. Incorporating this prior knowledge into the learning architecture, as a geometric prior, e.g., by adopting a graph \cite{velivckovic2017graph} or convolutional \cite{krizhevsky2012imagenet} neural network,  has been shown to improve both trainablity and generalization performance.


Recently, the quantum machine learning (QML) community has also focused on introducing geometric priors into quantum models \cite{larocca2022group, meyer2022exploiting, ragone2022representation}. For example, quantum graph neural networks preserve permutations symmetries, making them suitable for learning quantum tasks with a graph structure \cite{verdon2019learning, verdon2019quantum, ai2022decompositional, mernyei2022equivariant}; whilst quantum convolutional neural networks \cite{nguyen2022theory} preserve translations symmetry. As their classical counterparts, symmetry-preserving quantum models have the potential not only of reducing sample complexity, but also to mitigate quantum computing-specific issues such as barren plateaus \cite{schatzki2022theoretical,pesah2021absence}.

{\color{black}{A \emph{quantum recurrent neural network} (QRNN) architecture that can successfully learn temporal dependencies using a repeat-until-success mechanism was first proposed in \cite{bausch2020recurrent}. }} A separate QRNN model was later proposed in \cite{takaki2021learning}.
As illustrated in Fig. \ref{qrnn}, a QRNN applies the same parametrized quantum circuit (PQC) sequentially over time, whilst retaining historical information in unmeasured ``memory'' qubits. The model has been shown to be capable of learning sequential data \cite{takaki2021learning}, and its capacity was analyzed in \cite{elliott2022quantum} in comparison to counterpart classical models with  the same number of memory units. In particular, it was shown in \cite{elliott2020extreme,yang2021provable} that quantum models can describe temporal sequences with reduced memory, compared to their classical counterparts. 
A model with classical memory, was introduced in \cite{chen2020temporal} by integrating quantum models within a classical LSTM architecture. The latter model was also leveraged in \cite{wellsfargo} to define a reservoir computing solution with fixed dynamics. RNNs have also been integrated with quantum dynamical models in order to simulate non-Markovian dynamics in open system \cite{banchi2018modelling}. 

To the best of our knowledge, quantum models for temporal data processing have not yet been studied from the perspective of symmetry preservation. In this paper, we address this knowledge gap by focusing on models with quantum memory. To this end, we take the QRNN model in \cite{takaki2021learning} as foundation due to its efficiency, although the general approach presented here could be extended to any quantum model with a recurrent structure.

\subsection{Contribution}
{\color{black}{Inspired by the connection between gating and time warping unveiled in \cite{tallec2018can}, in this paper we investigate the problem of preserving symmetries caused by time-warping transformations in quantum dynamic models.}} The main contributions are as follows.
\begin{itemize}
    \item  We introduce the class of \emph{time warping-invariant QRNNs (TWI-QRNNs)}, which augment QRNNs with a \emph{quantum-classical adaptive gating} mechanism that chooses whether to apply a parameterized
unitary transformation or not at each time step as a function of the past samples of the input sequence
via a classical recurrent model.
\item The TWI-QRNN model class is derived from first principles  via a postulate of invariance to time warping. 
\item While the TWI-QRNN model class implements deterministic mappings, we also introduce a time warping-invariant stochastic model referred to as  \emph{time warping-invariant stochastic QRNNs (TWI-SQRNNs)}.
\item The capacity of TWI-QRNN and TWI-SQRNN model classes to successfully address time-warping transformations is experimentally demonstrated on examples with classical or quantum dynamics.
\end{itemize}


\section{Quantum Recurrent Neural Networks} \label{sec:QRNN1}
Formally, we study the problem of mapping classical sequences $x_{1:T}$ comprising $T$ samples $x_t$ with $t=1,2,...,T$,  to corresponding classical sequences $z_{1:T}$, comprising samples $z_t$ with $t=1,2,...,T$,  through a parameterized causal mapping operating over time $t$. {\color{black} {The QRNN model in \cite{takaki2021learning} implements a deterministic mapping between  sequence $x_{1:T}$ and sequence $z_{1:T}$ based on a parameterized quantum circuit (PQC) for the case in which both $x_t$ and $z_t$ are real valued.}} We will specifically focus here on settings in which the target samples $z_t$ are scalar, while input samples $x_t$ may be  arbitrary real vectors. In the following, we review the QRNN model, as well as the corresponding training criterion. Furthermore, we point out a connection between QRNNs and dissipative quantum neural networks (QNNs) \cite{sharma2022trainability}\cite{Heimann2022LearningCO}, which will be leveraged in the following section to introduce the proposed model class.

\subsection{Quantum Recurrent Neural Networks}

\begin{figure*}[tbp]
\centering
\includegraphics[width=0.9\textwidth]{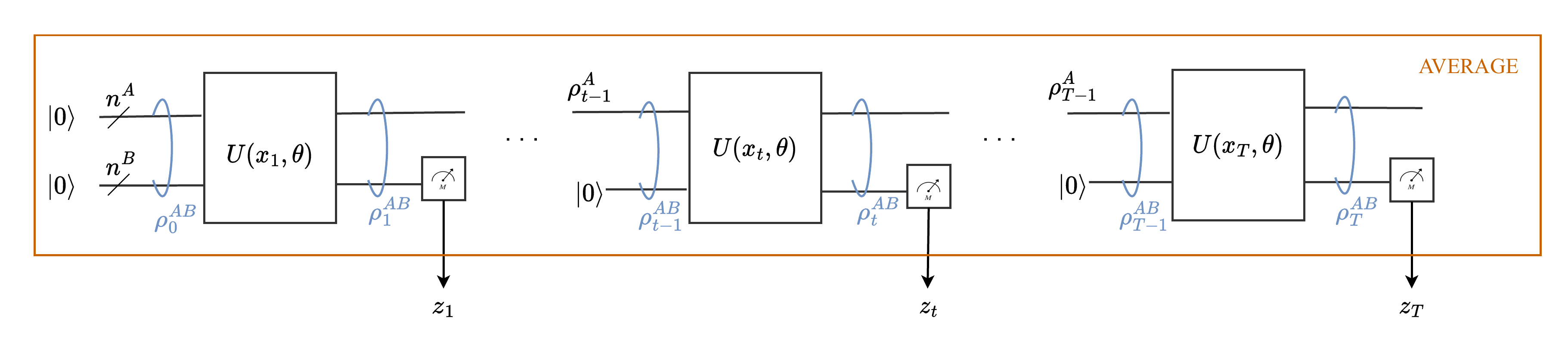}
\caption{Illustration of the QRNN model introduced in \cite{takaki2021learning}. 
}
\vspace{-0.5cm}
\label{qrnn}
\end{figure*}

A QRNN is defined by a parameterized unitary transformation $U(x, \theta)$ operating on a register of  $n=n^A+n^B$ qubits. The unitary is a function of the real-valued input vector $x$ and of a vector of real-valued model parameters $\theta$. {\color{black}{The ansatz, i.e., architecture, assumed in \cite{takaki2021learning} for the unitary $U(x,\theta)$ includes an input encoding layer with single-qubit rotations dependent on input $x$, followed by parameterized processing layers with both single-qubit and two-qubit gates. Note, however, that we allow the dependence of the unitary $U(x,\theta)$ to be arbitrary.}} {\color{black}{Therefore, the model accounts also for special input-encoding methods such as the repeat-until-success neural architecture from \cite{bausch2020recurrent} (see also \cite{cao2017quantum}).  }}

As illustrated in Fig. \ref{qrnn}, the unitary is applied at each time step by re-initializing the subregister of $n^B$ qubits to the ground state $|0\rangle$. After the application of the unitary to all $n$ qubits, the subregister of $n^B$ qbits is measured. Since the subregister of $n^A$ qubits is not measured, it can be used to propagate information from one time step to the next. Therefore, we will refer to the $n^A$-qubit subregister as the \emph{memory subregister}, while the subsystem of $n^B$ qubits is referred to as the \emph{output subregister}.


To elaborate, the density matrix describing the state of the register of $n$ qubits before the first application of the unitary is given by $\rho^{AB}_0=|0\rangle \langle 0|^A \otimes |0\rangle \langle 0|^B$, where the first state is defined on the memory subregister of $n^A$ qubits, and the second on the output subregister of $n^B$ qubits as described by the superscripts $A$ and $B$. {\color{black}{Note that the memory subregister is also intialized to $|0\rangle$.}} The application of the unitary $U(x_1,\theta)$, which encodes the first input sample $x_1$, yields the output state density $\rho_1^{AB}=U(x_1, \theta) \rho^{AB}_0 U(x_1, \theta)^{\dagger}$. Then, a measurement is made on the output subregister. 

The measurement  is  defined by a local observable $O^B$, which can always be expressed in a diagonal form 
$O^B = \sum_\lambda o_\lambda \ket{\lambda}\!\bra{\lambda}^B$ in a suitable basis $\ket{\lambda}^B$. As a result of a measurement with outcome $o_\lambda$, the wavefunction of the output subregister collapses onto the state $\ket{\lambda}^B$; and the density matrix of the memory subregister is changed to the reduced density \begin{equation}\rho^{A|\lambda}_1=\frac{(I^A\otimes {}^B\bra{\lambda})\rho^{AB}_1(I^A \otimes \ket{\lambda}^B)}{ {}^B\bra{\lambda}\rho^{B}_1\ket{\lambda}^B},\end{equation} where we have $\rho^B_1 = \mathrm{Tr}_A(\rho^{AB}_1)$ and $\mathrm{Tr}_X(\cdot)$ indicates the partial trace with respect to subregister $X\in\{A,B\}$. In the QRNN model, the measurement is repeated many times in order to obtain an estimate of the expectation value $\langle O^B\rangle$ of the observable $O^B$. After averaging over the measurement outcomes, the resulting density matrix of the memory subregister is given by $\rho^A_1 = \mathrm{Tr}_B(\rho^{AB}_1)$.
Note that this description of the state of the memory subregister is independent of the specific observable $O^B$, and is sufficient to predict the expectations of downstream measurements \cite{takaki2021learning}. Furthermore, we point the reader to Sec.~\ref{sec:stochastic}, which  studies a stochastic variant of the QRNN model in which a single measurement outcome $o_\lambda$ is evaluated at each time step.  

Generalizing to any time $t=1,...,T$, the memory subregister evolves according to the update rules\begin{equation}\label{eq:qrnn1}
\rho^{AB}_t =U(x_t, \theta) (\rho^A_{t-1} \otimes |0\rangle \langle 0|^B) U(x_t, \theta)^\dagger
\end{equation} and 
\begin{equation}\label{eq:qrnn2}
\rho^A_t =\mathrm{Tr}_B(\rho^{AB}_t),
\end{equation} with initialization $\rho^A_0=|0\rangle \langle 0|^A$. Furthermore, the output at each time $t$ is given by a function \begin{align}\label{prep_det}
    z_t = g(\langle O^B \rangle_t)
\end{align}of the expected value \begin{align}\label{expO}   \langle O^B \rangle_t = \text{Tr}[\rho^{AB}_t (I^A  \otimes O^B)] = \text{Tr}[\rho^B_t O^B]
\end{align}of the local  observable $O^B$, where $\rho^B_t=\text{Tr}_A[\rho^{AB}_t]$.

As mentioned earlier in this section and illustrated in Fig. \ref{qrnn}, evaluating the sequence of expected values $z_t$ requires running the circuit multiple times in order to approximately evaluate the expectations in (\ref{expO}) via empirical averages of the measurement outputs at all time steps $t=1,...,T$. 

\subsection{Training QRNNs}

Given a data set of input-output example pairs $(x_{1:T},\bar{z}_{1:T})$, training of a QRNN targets the minimization of the training loss. To this end, for each example pair $(x_{1:T},\bar{z}_{1:T})$, we define the quadratic training loss \cite{takaki2021learning}
\begin{align}\label{eq:loss}
    \ell (z_{1:T}, \bar{z}_{1:T}) = \frac{1}{T} \sum_{t=1}^T (\bar{z}_t - z_t)^2,
\end{align}
with $z_t$ is a function of the model parameter vector $\theta$ through  \eqref{prep_det}. The training loss (\ref{eq:loss}) is averaged over all examples in the training set, and it can be minimized via zeroth-order optimization schemes, such as the parameter shift rule \cite{schuld2021machine,banchi2021measuring}.

\subsection{QRNNs as Dissipative QNNs}\label{sec:diss}

In this subsection, we  observe a useful connection between QRNNs and dissipative QNNs \cite{sharma2022trainability,Heimann2022LearningCO}, a popular model for PQCs that mimics the operation of classical multi-layer perceptrons. We specifically focus on single-layer QRNNs as defined in \cite[Fig.~1]{sharma2022trainability}. A (single-layer) dissipative QNN encompasses  an input subregister and a number of output subregisters. A new output subregister is introduced at each processing step $t$, and a parameterized unitary $U_t$, also referred to as ``perceptron'', is applied to the input subregister and to the new output subregister. All output subregisters are finally measured. We note that dissipative QNNs are also related to quantum collision models used in quantum mechanics for the study of open quantum systems \cite{ciccarello2022quantum}.

\begin{figure*}[tbp]
\centering
\includegraphics[width=0.7\textwidth]{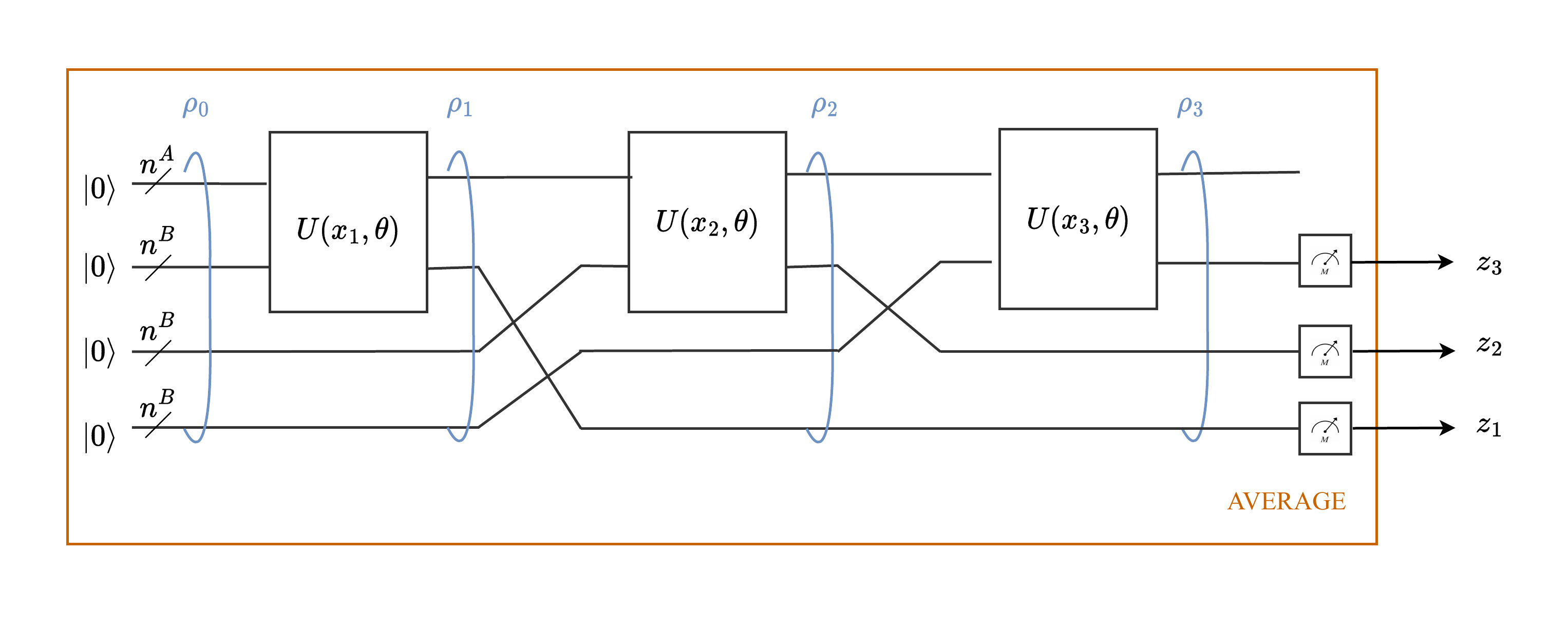}\vspace{-0.4cm}
\caption{The QRNN model in Fig. \ref{qrnn} can be interpreted as a variant of a dissipative QRNN \cite{sharma2022trainability,Heimann2022LearningCO}, as shown in the figure with $T=3$, in which the same parameter vector $\theta$ is reused by unitaries $U(x,\theta)$, and different data samples $x_t$ are loaded at each time $t$.}
\vspace{-0.5cm}
\label{cqrnn}
\end{figure*}

A dissipative QNN with  $T=3$ steps is  illustrated in Fig. \ref{cqrnn}.  Note that, unlike the dissipative QRNN model in \cite{sharma2022trainability,Heimann2022LearningCO}, in Fig. \ref{cqrnn}, a different input data sample $x_t$ is loaded at each step $t$, and the unitaries $U(x_t, \theta)$ used across steps $t$ share the same parameter vector $\theta$. It can be readily checked that the dissipative QNN in Fig.  \ref{cqrnn} is equivalent to the QRNN model in Fig. \ref{qrnn} if we identify the input subregister of the dissipative QNN with the QRNN's memory subregister  and the output subregisters with the QRNN's output subregister. This is in the sense that both models provide statistically equivalent outputs. 

{\color{black}{The main advantage of formulating a QRNN in terms of a dissipative QRNN is that the evolution of the system and the outputs given by (\ref{eq:qrnn1})-(\ref{expO}) can be more directly expressed without explicitly resorting to partial trace operations and intermediate measurements at each time step $t$
}}.  To this end, define a system, as exemplified in Fig.  \ref{cqrnn} with $n_T=n_A+T n_B$ qubits, corresponding to one memory subregister  of $n^A$ qubits and $T$ output subregisters of $n^B$ qubits. The output subregisters are indexed as $t=1,2,...,T$ as in Fig. \ref{cqrnn}. Initializing the density state of such system to the ground state $\rho_0=|0\rangle \langle 0|$, where $|0\rangle$ is a separable ground state across the $n_T$ qubits, the density state evolves as \begin{align}\label{eq:cq1}
    \rho_{t+1} &= V_t(x_t, \theta) \rho_t V_t(x_t, \theta)^{\dagger},
\end{align}where $V_t(x_t, \theta)$ is a unitary matrix that applies unitary $U(x_t,\theta)$ to the memory subregister (first wire in Fig. \ref{cqrnn}) and to the  $t$-th output subregister, while identity operators are applied to all other subregisters. Furthermore, the outputs (\ref{expO}) can be written as\begin{equation}\label{eq:cq2}   \langle O^B \rangle_t = \text{Tr}[\rho_t O_t^B],
\end{equation}where  the local observable $O_t^B$ applies only to the $t$-th memory subregister.

\section{Time Warping-Invariant QRNNs}

In this section, we introduce the proposed TWI-QRNN model class. We start by defining the concept of time warping-invariance. Then, we derive the TWI-QRNN  model, which is finally described, along with the corresponding training problem.

\subsection{Time Warping-Invariance}
Following \cite{tallec2018can} and \cite{bronstein2021geometric}, in order to define time warping-invariance, we consider a continuous-time formulation in which the time axis is defined by a real variable $t$.  Round parentheses will be used to denote dependence on the continuous time $t$. Furthermore, we will take $a(t)$ to represent either the value of the function $a(\cdot)$ at time $t$, or the entire function $a(\cdot)$, as it will be clear from the context. 

To start, we define a \emph{time warping operation} $c(t)$ as any monotonically increasing function of time. Let us fix  some subset $\mathcal{C}$ of time warping operations. An example is given by the \emph{linear time warping} family $\mathcal{C}$ given by operations of the form $c(t)=a t$ for some range of values $a>0$, which correspond to a stretching ($a<1$) or shrinking ($a>1$) of the time axis. Time warping-invariance is a property of a parameterized model class $z(t)=f(x(t),\theta)$ of mappings between an input continuous-time signal  $x(t)$ and an output continuous-time signal $z(t)$. Note that the mapping is arbitrary, and not restricted to memoryless functions. Furthermore, a mapping in the class is identified by the model parameter $\theta$. 

Model class $f(\cdot,\theta)$ is said to be \emph{time warping-invariant} with respect to the family $\mathcal{C}$, if for any $z(t)=f(x(t),\theta)$ produced by the model with some model parameters $\theta$ given input $x(t)$, there exist model parameters $\theta'$ that yield the output $z(c(t))=f(x(c(t)),\theta')$ given input  $x(c(t))$ for any function $c(t)$ in $\mathcal{C}$. In words, the model class can reproduce time-warped versions of its input-output pairs.

It is emphasized that the definition of time warping-invariance given above applies only to continuous-time processes. Therefore, its extension to discrete-time processes, first introduced in \cite{tallec2018can}, requires the use of discrete-time approximations. As we detail in the next subsection, these approximations entail that the notion of time warping-invariance is more loosely defined as compared to the ``spatial'' invariances studied in quantum geometric machine learning \cite{ragone2022representation}. We may hence refer to the resulting invariance property as a \emph{quasi}-invariance as in \cite{tallec2018can}, but we will not make this distinction explicit in the following.

\subsection{Time Warping-Invariance for QRNNs}\label{sec:twi_qrnn}

The time warping-invariance
property defined in the previous subsection can be applied to QRNNs
as long as one specifies a suitable continuous-time extension of the
defining equalities (\ref{eq:qrnn1})-(\ref{expO}), or equivalently
(\ref{eq:cq1})-(\ref{eq:cq2}). Here, we adopt a stronger definition
of time warping-invariance based on the formulation
(\ref{eq:cq1})-(\ref{eq:cq2}), whereby the invariance condition is
imposed on the input-output pairs $(x_t,\rho_t)$. Note that, if
invariance holds in terms of the density matrices $\rho_t$, then it
also holds, a fortiori, on the expected values (\ref{eq:cq2})
producing the output samples $z_t$. In the rest of this subsection,
we provide a derivation of the TWI-QRNN model that differs from 
the steps of classical models 
\cite{tallec2018can},~\cite{bronstein2021geometric} and is more 
suitable for a quantum mechanical description of the problem. A direct extension of the arguments in \cite{tallec2018can},~\cite{bronstein2021geometric} can be found in the Appendix.
 The reader only interested in the
definition of the model can move on directly to the next subsection. 

Let us start from the definition \eqref{eq:cq1}
and apply a time-warping operation so that the input signal observed by the QRNN at time $t$ is given by $x(c(t))$. Denote as $\rho(t)$ a continuous-time version of the density matrix $\rho_t$ obtained with the original input $x(t)$, and as $\rho(c(t))$ the corresponding  density matrix obtained with the time-warped input $x(c(t))$ under time warping-invariance. 
The change in density matrix from one discrete-time step $t$ to the next, $t+1$, 
can be related to the derivatives with respect to the warped time axis as
\begin{equation}
	\rho(c(t+1))-\rho(c(t)) = \int_{t}^{t+1} 
	\frac{d\rho(c(t))}{d c(t)} \frac{d c(t)}{dt} \,dt,
	\label{eq:integralid}
\end{equation}
where we have used the definition of integral and 
changed the integration variable from $c(t)$ to $t$. 

To proceed, we make the approximation of considering the time-warping derivative $d c(t)/dt$ to be constant 
in the interval $[t,t+1]$, so that it can be taken out 
of the integral in (\ref{eq:integralid}). 
Furthermore, in order to account for the desired time warping-invariance of the output density matrix, we assume that the continuous-time
evolution operator is approximately invariant under time-warping 
transformations, i.e., 
\begin{equation}
	\frac{d\rho(c(t))}{d c(t)} \approx \frac{d\rho(t)}{dt}.
	\label{eq:scaleinvariance}
\end{equation}
With regards to the approximations put forth in this paragraph, we note that we are only interested 
in the evolution of the density at the end of the discrete time-step from time $t$ to $t+1$ as per \eqref{eq:cq1}, and hence we can
ignore constraints on the underlying infinitesimal physical evolution as long as they are consistent with the discrete-time evolution. {\color{black}{Importantly, we emphasize that the approximate equality \eqref{eq:scaleinvariance} is not intended to indicate that a meaningful continuous-time physical process $\rho(t)$ exists that satisfies \eqref{eq:scaleinvariance} with a strict equality. Rather, the continuous-time quantum process is just used as a motivational starting point for the introduction of the TWI-QRNN model. For clarity, and to acknowledge the approximations made, we will use $\tilde{\rho}(\cdot)$ in the rest of the derivation. }}

Inserting \eqref{eq:scaleinvariance} into \eqref{eq:integralid}
and using the above approximations, we obtain 
\begin{align}
	&\tilde{\rho}(c({t+1}))-\tilde{\rho}(c(t))  \approx \frac{d c(t)}{dt} 
	\int_{t}^{t+1} 
	\frac{d\tilde{\rho}(t)}{d t} \,dt \nonumber\\
    &= 
	\frac{d c(t)}{dt} (\tilde{\rho}({t+1})-\tilde{\rho}(t)).
	\label{eq:approxrho2}
\end{align} 
In (\ref{eq:approxrho2}), the density $\tilde{\rho}(t+1)$ is given by the output of the 
QRNN, via ~\eqref{eq:cq1}, with previous state
$\tilde{\rho}(c(t))\approx \tilde{\rho}(t)$. Moreover, 
from the point of view of a discrete QRNN model under time warping, the time $c(t)$ corresponds to discrete time $t$, the value $x(c(t))$ to sample $x_t$, and density matrix $\tilde{\rho}(c(t))$ to $\tilde{\rho}_t$.  
Using these definitions, 
we get the approximate equality 
\begin{align}\label{lQRNN}
    \tilde{\rho}_{t + 1} \approx \left(1-\frac{\textrm{d} c(t)}{\textrm{d} t}\right)\tilde{\rho}_{t} + \frac{\textrm{d} c(t)}{\textrm{d} t} V_t(x_t, \theta) \tilde{\rho}_t V_t(x_t, \theta)^{\dagger}.
\end{align} 
{\color{black}{We observe that the right-hand side in \eqref{lQRNN} is a valid density matrix if the inequality $\textrm{d}c(t)/\textrm{d}t \leq 1$ holds. The same limitation applies to the derivation presented in \cite{tallec2018can} for classical models (see also Appendix), and it stems from approximations done in going from discrete to continuous time and back.}}

In other words, under condition (\ref{lQRNN}), the density matrix $\tilde{\rho}_{t+1}$ is produced by leaving the previous density matrix $\tilde{\rho}_{t}$ unchanged with probability $(1-\textrm{d} c(t)/\textrm{d} t)$ and by applying the unitary $V_t(x_t, \theta)$ otherwise.  The update (\ref{lQRNN}) is the key equation defining TWI-QRNNs, as it will be elaborated on in the next subsection.

\subsection{TWI-QRNNs}\label{sec:warpsub}
Based on the analysis in the previous subsection, we define the class of TWI-QRNNs using the partial trace formalism (\ref{eq:qrnn1})-(\ref{expO}) as follows. A TWI-QRNNs applies the updates
\begin{align}\label{eq:qrnnmodel}
&\rho^{AB}_t =(1-\alpha_t) (\rho^{A}_{t-1}\otimes |0\rangle \langle 0|^B)\nonumber\\
&+\alpha_t U(x_t, \theta) (\rho^A_{t-1} \otimes |0\rangle \langle 0|^B) U(x_t, \theta)^\dagger,
\end{align} where $\alpha_t\in [0,1]$ is a probability, along with the equalities (\ref{eq:qrnn2})-(\ref{expO}). Accordingly, when evaluating the output $z_t$ for time $t$, a TWI-QRNN generates a number of realizations of  $t-1$ Bernoulli random variables $b_1,...,b_{t-1}$, with each random variable $b_t \sim \text{Bern}(\alpha_t)$ being equal to $1$ with probability $\alpha_t$. For each realization $b_1,...,b_{t-1}$, at time $t'<t$, a unitary $U(x_{t'},\theta)$ is applied to the memory and output subregisters if $b_{t'}=1$ and an identity is applied otherwise, followed by a measurement of the output register. The outputs of the measurement of the output register at time $t$ are then averaged to obtain $z_t$.

By the analysis in the previous subsection, culminating in (\ref{lQRNN}), the probability $\alpha_t$ should be chosen equal to the time-warping derivative $\mathrm{d}c(t)/\mathrm{d}t$. To gain insight on this selection, let us consider the case of a linear time warping family $\mathcal{C}$ defined as $c(t)=a t$ with $a<1$, which corresponds to a stretching the inputs and outputs by a factor of $1/a$. Accordingly, by the definition of time warping-invariance, if the input $x_t$ is stretched by a factor of $1/a$, the model class should be able to reproduce an equally stretched density output sequence $\rho_t$. Therefore, assuming $1/a$ to be an integer, if each input sample $x_t$ of the original sequence is kept constant for $1/a$ consecutive samples,  the output samples $\rho_t$ of the output sequence should also be constant for consecutive intervals of $1/a$ samples. The update (\ref{eq:qrnnmodel}) implements a probabilistic version of this operation, whereby each sample $\rho_t$ is held constant, on average, for $1/(1-\alpha_t)$ consecutive samples. Therefore, setting $\alpha_t =1- a$ approximately satisfies time warping-invariance with respect to the family $\mathcal{C}$.

We finally observe that the mechanism described in this subsection, which is based on $t-1$ classical binary variables and switches,  could be readily replaced by a fully quantum implementation with controlled-$U$ gates and $t-1$ of controlling ancilla qubits. However, this alternative architecture is rather inefficient for sequences containing a sufficiently large number of time samples.

\subsection{Adaptive Gating Mechanism}
\begin{figure*}[tbp]
\centering
\includegraphics[width=0.9\textwidth]{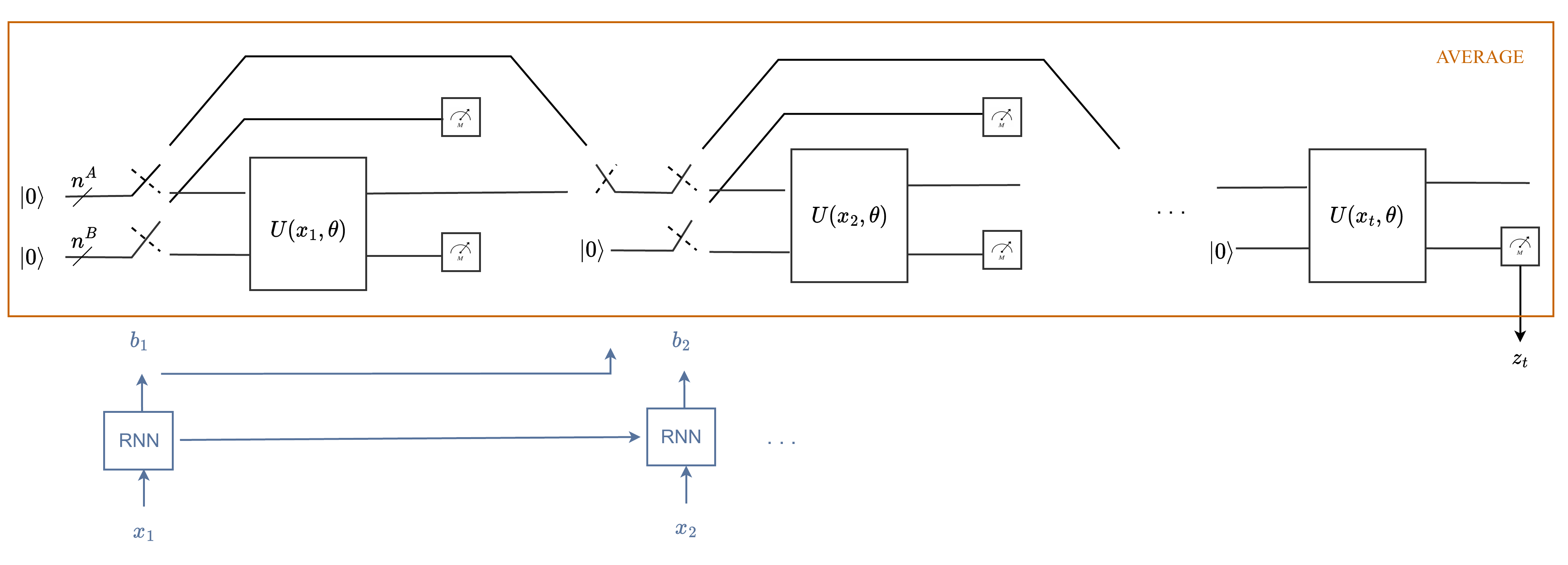}\vspace{-0.4cm}
\caption{Illustration of the operation of the TWI-QRNN model.}
\vspace{-0.5cm}
\label{cqrnn copy}
\end{figure*}

{\color{black}{Whilst the motivational derivation above suggests the choice of $\alpha$ as $\textrm{d} c(t)/\textrm{d} t$, the proposed algorithm practically replaces the unknown derivative in the model (\ref{eq:qrnnmodel}) with a classical recurrent neural network model (RNNs) that infers a suitable time-warping probability $\alpha_t$ from the input data sequence in a causal fashion.}} Specifically, we write the probability $\alpha_t$ as 
\begin{align}\label{derivative}
    \alpha_t = \sigma (\phi_t),
\end{align}
{\color{black}{where $\sigma(a)=(1+\exp(-a))^{-1}$ denotes a logistic sigmoid}} with \emph{hyperparameter vector} $\phi_t$ being the output of a model defined by the updates
\begin{align}\label{phi_hyper}
    \phi_t = W_x x_t + W_h h_t + b,
\end{align}
and
\begin{align}
\label{phi_hyper1}
    h_t = \text{tanh} (W^h_x x_t + W^h_h h_{t-1} + b),
\end{align}
where $W = \{W_x, W_h, W^h_x, W^h_h\}$ denote learnable parameters. The overall TWI-QRNN model is shown in Fig.~\ref{cqrnn copy}. 

{\color{black}{We emphasize that the classical RNN model is not required if one has prior knowledge of the time-warping function. Furthermore, we intentionally  avoid introducing gating mechanisms in the RNN. This choice is motivated by the designated role of the  RNN as a mechanism to estimate the time-warping function, with the model memory being managed by the quantum system via the described gating scheme. That said, we note that more complex tasks may call for more sophisticated gating mechanisms across both quantum and classical models.}}

\subsection{Training TWI-QRNNs}

Let us define as 
\begin{align}\label{Bern}
    q(b_t \, | \,x_{1:t}, W) = \sigma (\phi_t)^{b_t} (1-\sigma (\phi_t))^{1-b_t}
\end{align}the variational distribution of the Bernoulli variable $b_t$ dictating whether or not the current unitary $U(x_t,\theta)$ is applied to the input and memory subregisters as per (\ref{eq:qrnnmodel}). The joint distribution of the variables $b_{1:T}$ is then given as $q(b_{1:T}|x_{1:T},W)=\prod_{t=1}^{T} q(b_t \, | \,x_{1:t}, W)$. Note that the samples $b_{1:T}$ are conditionally independent given the input sequence $x_{1:T}$, and that the variable $b_t$ depends causally on the input samples $x_{1:T}$. Training a TWI-QRNN amounts to the optimization of the  variational parameters $W$ and of the PQC parameters $\theta$.

To specify the training problem, we define the loss for each example $(x_{1:T},\bar{z}_{1:T})$ as the expectation
$\E [\ell (z_{1:T}, \bar{z}_{1:T})]$ of the loss function (\ref{eq:loss}), where the average is taken with respect to the random variables $b_{1:T}$, which, in turn, determine the random outputs $z_{1:T}$. To estimate this expectation, one can leverage realizations of the outputs $z_{1:T}$ obtained as described in the previous subsection by drawing samples from the variational distribution $q(b_{1:T}|x_{1:T},W)$.  

The resulting training loss is minimized via gradient descent over $W$. Specifically, the variational parameters $W$ are optimized using the log-derivative trick \cite{williams1992simple} to estimate the gradient, whilst optimization over $\theta$ is carried out using zeroth-order optimization as for QRNNs.
    


\begin{figure*}[tbp]
\centering
\includegraphics[width=0.9\textwidth]{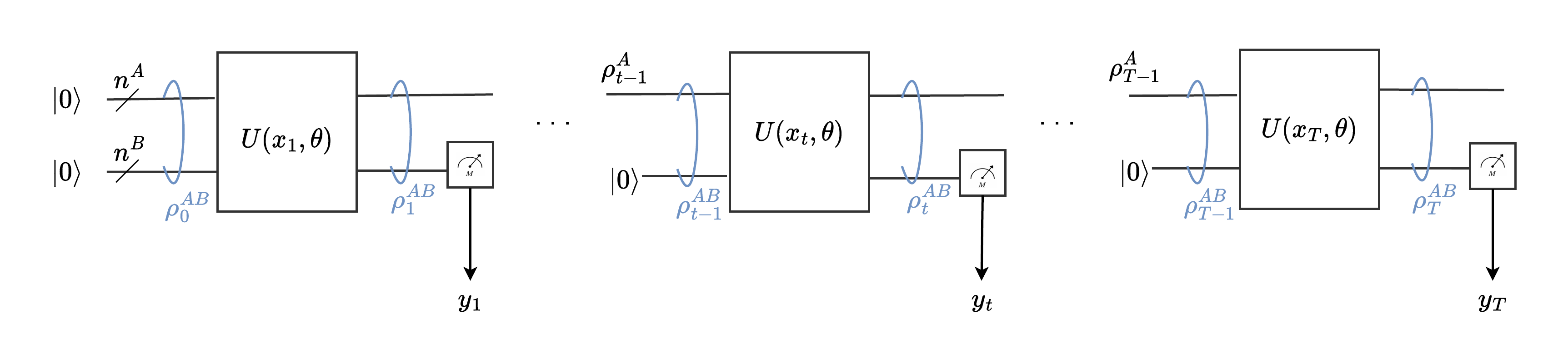}
\caption{Illustration of the proposed SQRNN model.}
\vspace{-0.5cm}
\label{sqrnn}
\end{figure*}

\section{Stochastic Time Warping-Invariant QRNNs}\label{sec:stochastic}

In this section, we introduce a probabilistic counterpart of the QRNN, referred to as stochastic QRNN (SQRNN), which is then extended to the time warping-invariant class TWI-SQRNN. 

\subsection{SQRNNs}

A SQRNN is a probabilistic model that maps an input real-valued vector sequence $x_{1:T}$ to a discrete-valued sequence $y_{1:T}$, with $y_t$ being a string of $n_B$ bits. Note that equivalently, we can consider the output sample $y_t$  to take one out of $2^{n^B}$ possible values. Accordingly, without loss of generality, we write $y_t \in \{0:2^{n^B}-1\}$. 
Unlike the QRNN model introduced in Sec. II, an SQRNN outputs sequence $y_{1:T}$ via a \emph{single} run of the circuit, hence not requiring an empirical average over multiple runs of the circuit. Specifically, as illustrated in Fig.~\ref{sqrnn}, the output $y_t$ is the random output of the measurement of the output subregister at time $t$. 

To elaborate, let us fix a projective measurement defined by projection matrices $\Pi^B_y$ with $y\in \{0:2^{n^B}-1\}$. The projection matrices act on the output subregister. Then, the probability of the outcome $y_t$ at time $t$ given the past and current samples $x_{1:t}$ of the input sequence is given by Born's rule as \begin{align}\label{like_Born}
    p (y_t \, | \, x_{1:t}, \theta) = \mathrm{Tr}(\rho^{AB}_t (I^A \otimes \Pi^B_{y_t})),
\end{align}
where $\Pi_{y_t}$ is applied only to the output subregister and the density matrix $\rho^{AB}_t$ is defined as in Sec. \ref{sec:QRNN1}. 

Given an example $(x_{1:T},\bar{y}_{1:T})$, the training loss is defined as the cross-entropy loss 
\begin{align}\label{loss}
    \ell (x_{1:T}, \bar{y}_{1:T}) &= - \frac{1}{T} \sum_{t=1}^T \log p (\bar{y}_t \, | \, x_{1:t}, \theta) \nonumber\\
    &= - \frac{1}{T} \sum_{t=1}^T \log \mathrm{Tr}(\rho^{AB}_t (I^A \otimes \Pi^B_{\bar{y}_t})).
\end{align}We observe that this loss is the negative logarithm of the product of conditional marginals. Its minimization may be interpreted as a form of pseudo  maximum likelihood \cite{solomon2017pseudo}. An alternative, but more complex, criterion would be the negative logarithm of the joint distribution of sequence $\bar{y}_{1:T}$, which can be obtained from the joint density $\rho_t$ described in Sec. \ref{sec:diss}. The training problem with loss (\ref{loss}) can be addressed using the same tools mentioned in Sec. \ref{sec:QRNN1} for QRNNs.

\subsection{TWI-SQRNNs}
In the previous section, we have introduced TWI-QRNNs by imposing time warping-invariance in terms of the density matrices produced by the equivalent dissipative QNN. Since the probabilities (\ref{like_Born}) have the same form as the expectations (\ref{expO}) defining the outputs of QRNNs, the same reasoning applies to SQRNNs $\rho^{AB}_t$. 

Based on this observation, we define TWI-SQRNNs in a manner analogous to TWI-QRNNs. The only caveat is that each output $y_t$ is obtained via a \emph{single} pass through the circuit up to time $t$, rather than by carrying out several runs through the circuit. Accordingly, when evaluating the output $y_t$ for time $t$, a TWI-SQRNN generates a single realization of  $t-1$ Bernoulli random variables $b_1,...,b_{t-1}$, with each random variable $b_t \sim \text{Bern}(\alpha_t)$ being equal to $1$ with probability $\alpha_t$ in (\ref{derivative}). For the given realization $b_1,...,b_{t-1}$, at time $t'<t$, a unitary $U(x_{t'},\theta)$ is applied to the memory and output subregisters if $b_{t'}=1$ and an identity is applied otherwise, followed by a measurement of the output register.

Finally, training is carried out in a manner analogous to TWI-QRNNs by defining the training loss as the expectation of the cross-entropy loss (\ref{loss}) with respect to the variational distribution $q(b_{1:T}|x_{1:T},W)$.

\section{Experiments}
In this section, we provide experimental results to elaborate on the performance of TWI-QRNNs -- and their stochastic counterpart, TWI-SQRNNs -- as opposed to conventional (s)QRNNs, in the presence of time-warping distortions of the input sequence. As a classical benchmark, we use an LSTM RNN, which by \cite{tallec2018can} also preserve invariance to time-warping via gating mechanisms.

\subsection{Learning Tasks}
\subsubsection{Remembering a Cosine Wave}

\begin{figure*}[tbp]
\centering
\subfigure{\includegraphics[width=0.3\textwidth]{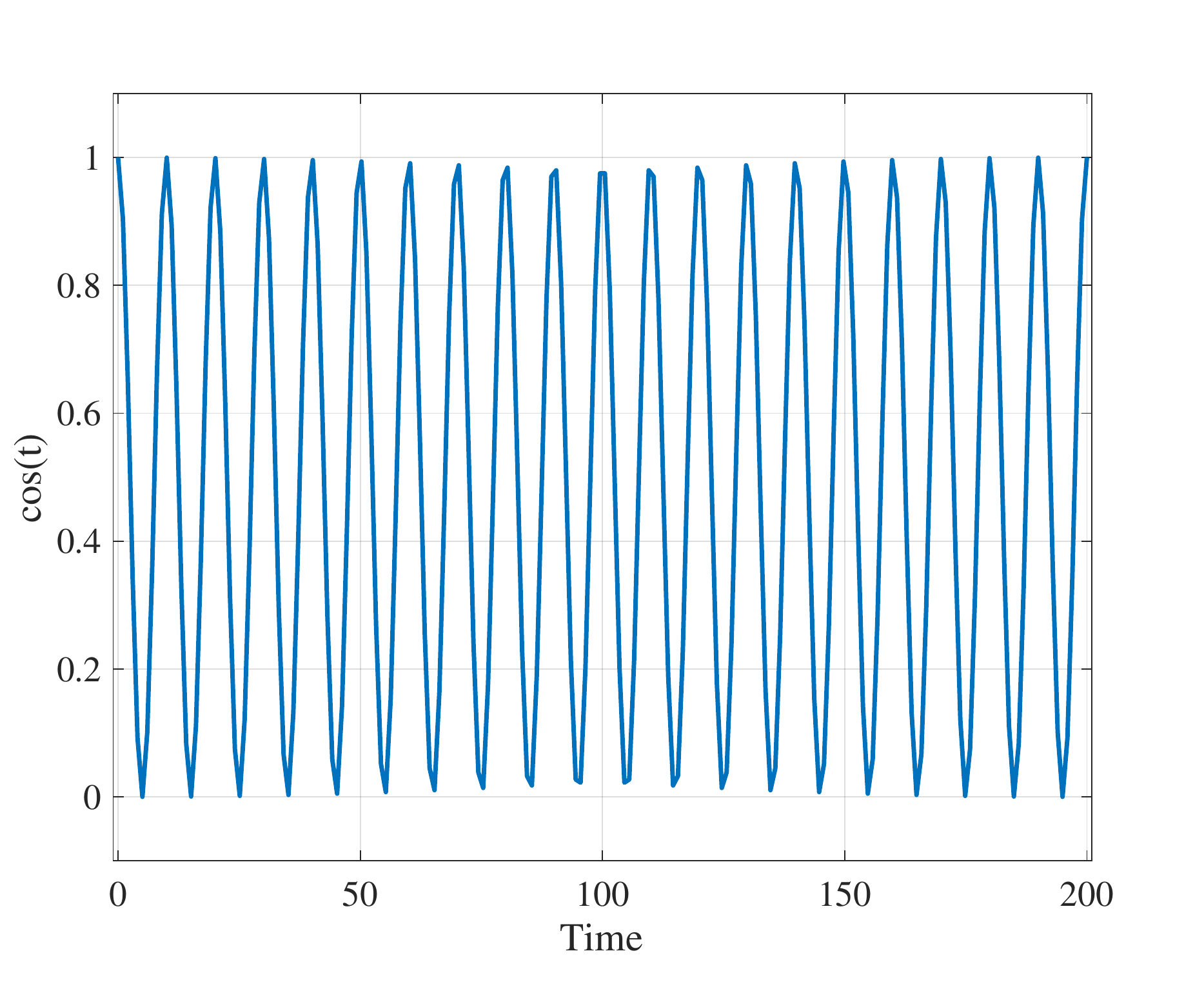}}
\subfigure{\includegraphics[width=0.3\textwidth]{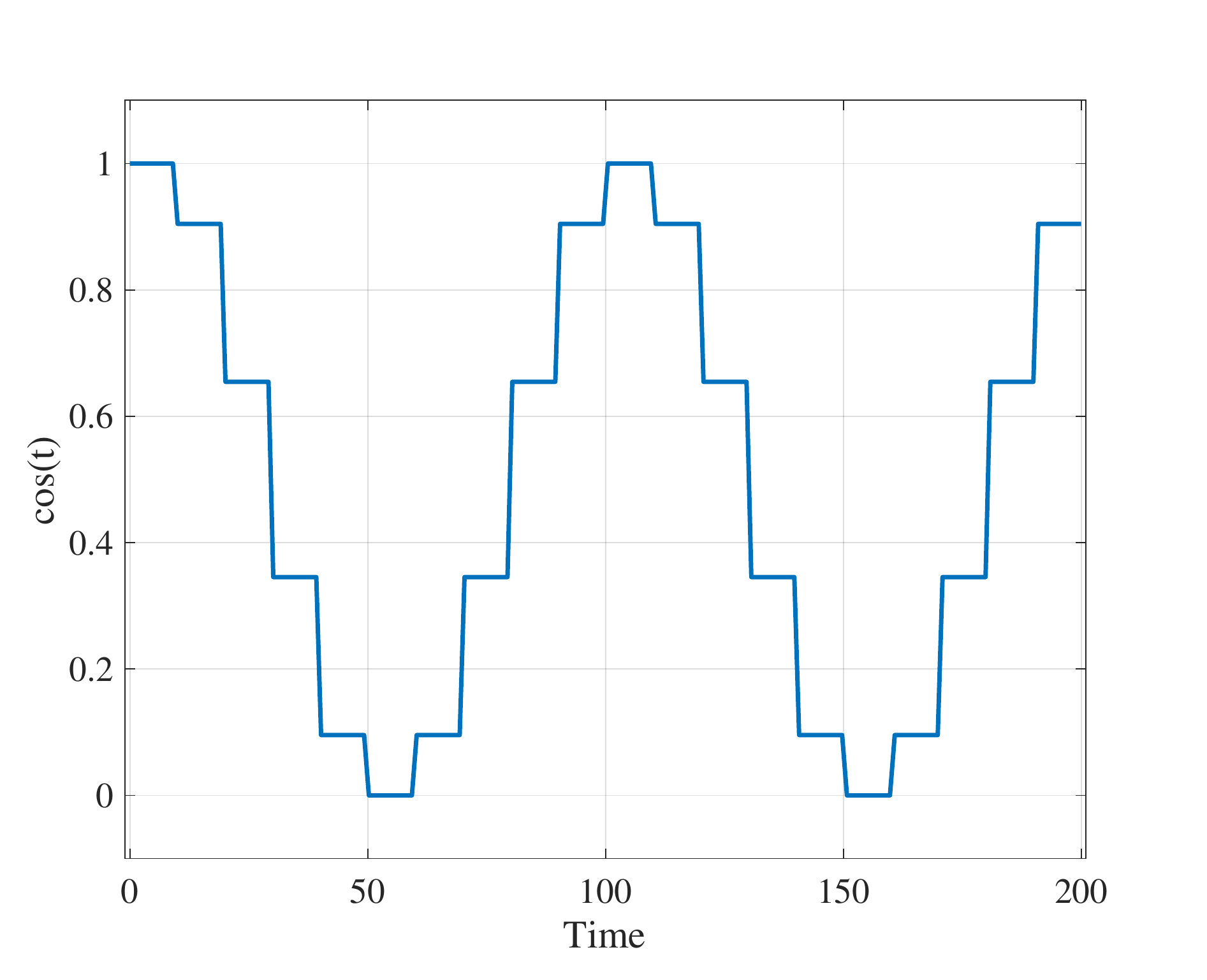}}
\subfigure{\includegraphics[width=0.3\textwidth]{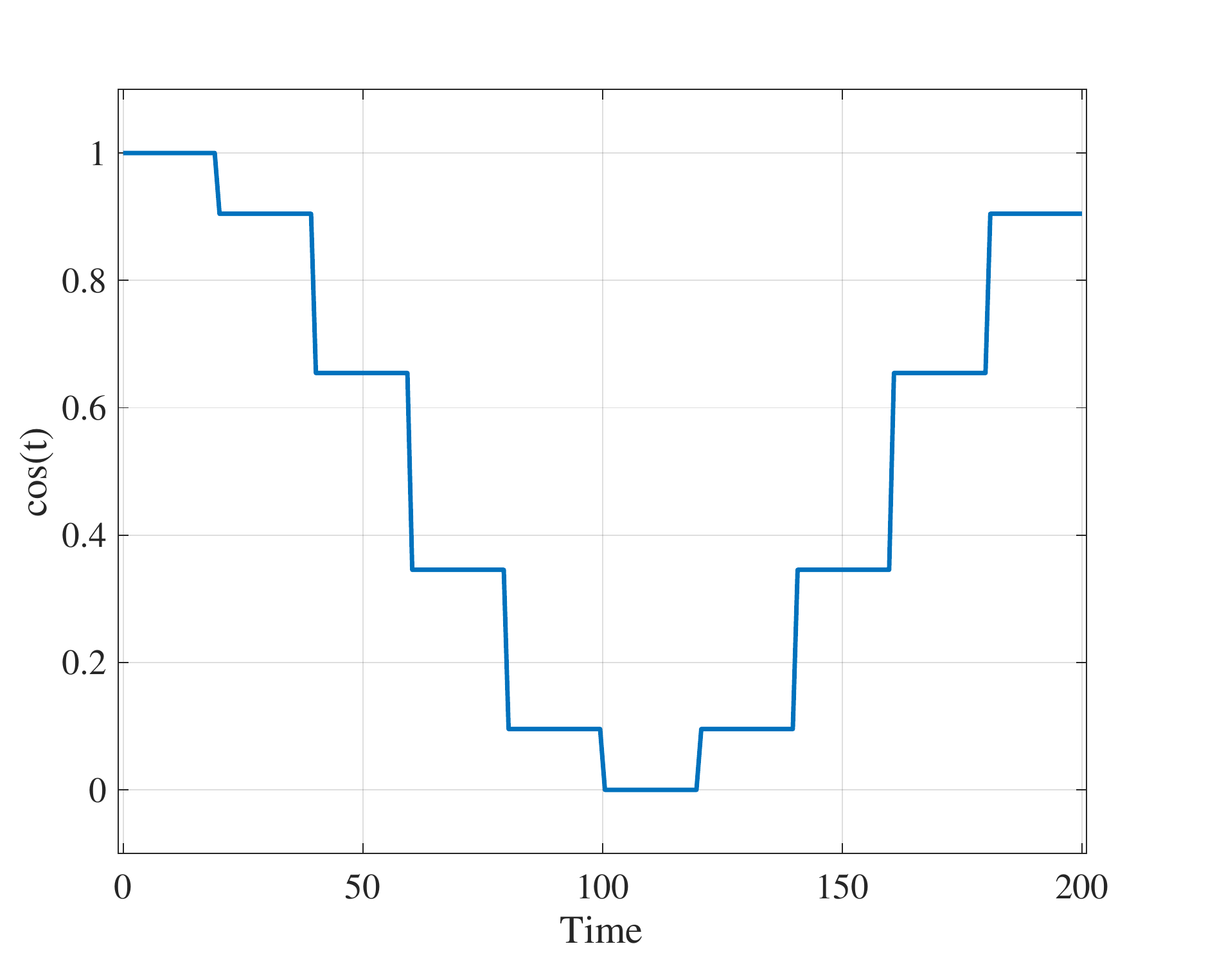}}
\caption{(left) The ``unwarped'' cosine function  (\ref{cos}) used for the first experiment (as in \cite{takaki2021learning}); and the corresponding (linearly) time-warped signals for $a = 0.1$ (middle) and $a = 0.05$ (right).}
\label{sin_eg}
\end{figure*}

To start, we consider the task of remembering the past sample of a simple classical time sequence $x_t$, namely a, possibly time-warped, discrete-time cosine function. Accordingly, we set the target sequence as $\bar{z}_{t}=x_{t-1}$ for (TWI-)QRNNs and $y_{t}=x_{t-1}$ for (TWI-)SQRNNs. For the case without time warping, reference \cite{takaki2021learning} demonstrated that the original QRNN model is capable of  successfully implementing this task. In contrast, here we study the robustness of the models under study to a time warping of the input sequence. 
To elaborate, the considered cosine function is given by
\begin{align}\label{cos}
    \text{cos}(t) = \frac{\text{cos}\left(\pi\frac{t}{5}\right)+1}{2}.
\end{align}
The ``unwarped'' sequence in \eqref{cos} is shown in Fig.~\ref{sin_eg}(left) for  $T = 200$ samples $x_t=\cos(t)$ obtained at time instants $t=1,2,...,T, \, T=200$. For the stochastic models, which produces discrete outputs, the value of the cosine in \eqref{cos} is discretized to $2^{n_B}$ levels equally spaced in the interval $[-1,1]$ with $n_B=3$ (see next subsection for further details on the model architecture).

We apply linear time-warping to obtain ``warped'' input sequences $x_t$ as discussed in Sec. \ref{sec:warpsub}. Accordingly, given the unwarped sequence $x_t$ obtained by sampling the original cosine signal $\cos(t)$, we repeat each sample for $1/a$ time instants, where  $a$ is such that the ratio $1/a$ is an integer. This setting accounts for a sampler with memory, as a new sample is observed every $1/a$ time steps. Examples of  warped sequences $x_t$ are shown in Fig.~\ref{sin_eg} for $a=0.1$ (middle) and $a=0.05$ (right).

\subsubsection{Predicting Spin Dynamics} 

Following \cite{takaki2021learning}, we study next the problem of predicting the expected value of an observable for the quantum spin dynamics of a three-qubit system. The density state $\sigma(t)$ of the three-qubit system evolves in continuous time according to the Lindblad master equation
\begin{align}\label{Lindblad}
    \frac{\textrm{d} \sigma(t)}{\textrm{d} t} &= -i [H, \sigma(t)] \nonumber\\
    &+ \frac{1}{2} \sum_i \left[2 C_k \sigma(t) C_k^{\dagger} - \sigma(t) C_k^{\dagger} C_k - C_k^{\dagger} C_k \sigma(t)\right],
\end{align}
with Hamiltonian
\begin{align}\label{Hamiltonian}
    H &= - \frac{1}{2} \sum_{i=1}^3 h_i Z_i \nonumber\\
    &- \frac{1}{2} \sum_{i=1}^2 \left(J_i X_i X_{i+1} + J_i Y_i Y_{i+1} + J_i Z_i Z_{i+1}\right),  
\end{align}
and matrices $C_k = c(X_k+ Y_k)$. The coefficients are set to $h_k = 2 \pi$, $J_k = 0.1 \pi$, and the dissipation rate equals $c = \sqrt{0.0002}$. The initial state is set to $\rho(0)=\ket{+}^{\otimes n}$, and the sequence is  sampled at continuous time instants $\Delta T,2\Delta T,...,200\Delta T$ with  $\Delta T=1/20$, which correspond to the discrete time instants $t=1,2,...,T$, respectively, with $T=200$. The unwarped input $x_t$ is then given by the corresponding expected values of the $Z$ observable for the first qubit of the three qubits.

\begin{figure*}[tbp]
\centering
\subfigure{\includegraphics[width=0.4\textwidth]{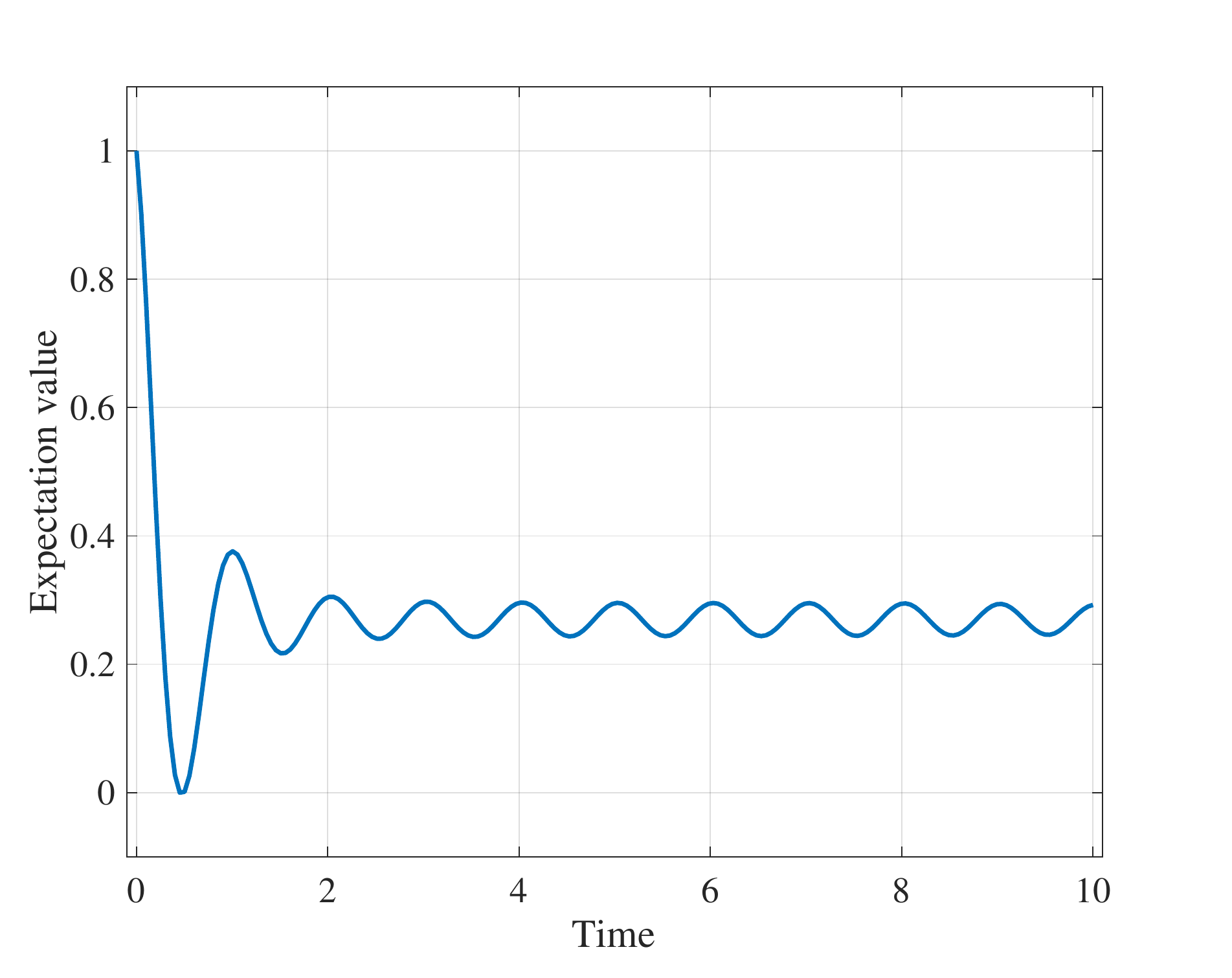}}
\subfigure{\includegraphics[width=0.4\textwidth]{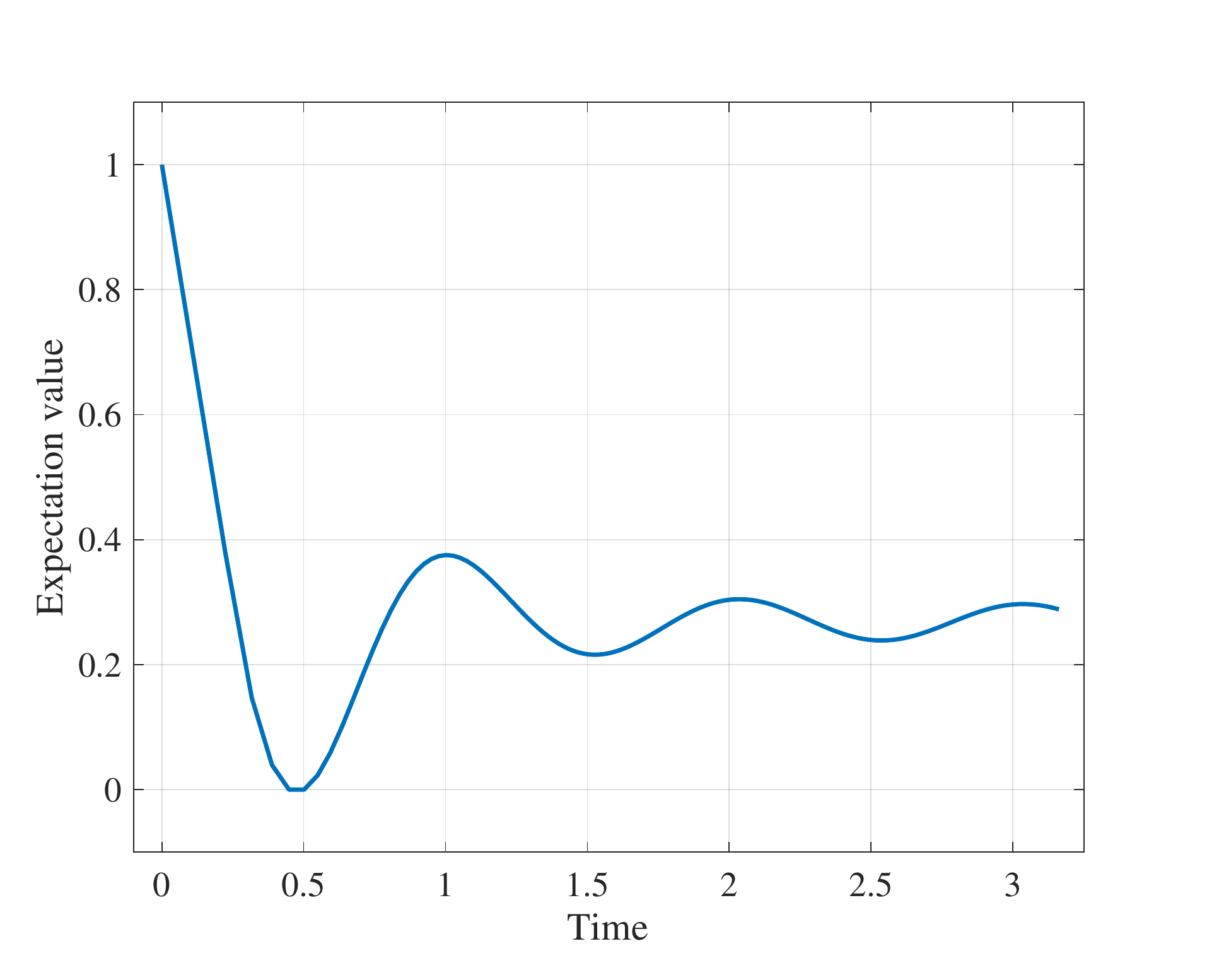}}
\caption{(left) The ``unwarped'' expected value of the observable for the three-qubit spin dynamics generated by \eqref{Lindblad}; and (right) the corresponding time-warped signal with non-linear transformation $c(t) = \sqrt{t}$.}
\label{fig:Lindblad}
\end{figure*}

Non-linear time-warped sequences are generated by setting $c(t) = \sqrt{t}$, and sampling the solution of the master equation (\ref{Hamiltonian}) at continuous time instants $\sqrt{\Delta T},\sqrt{2\Delta T},...,\sqrt{200\Delta T}$, which correspond to the discrete time instants $t=1,2,...,T=200$. This setting represents an instance in which the sampling interval varies over time.



\subsection{Architectures and Hyperparameters} 
\subsubsection{Circuit}
\begin{figure*}[tbp]
\centering
\includegraphics[width=0.7\textwidth]{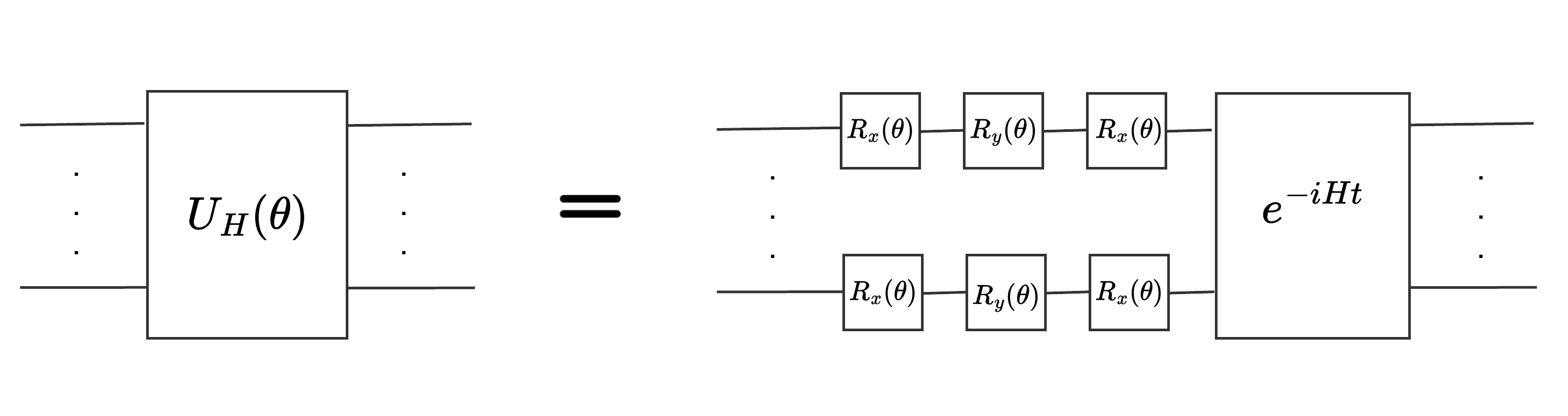}\vspace{-0.4cm}
\caption{The evolution part of the circuit utilized in the numerical simulations. The rotation angles
$\theta$ act as parameters of the circuit. The Hamiltonian $H$ is fixed and given by \eqref{Hamiltonian_circ}.}
\vspace{-0.5cm}
\label{UH}
\end{figure*}

Throughout this section, we adopt models with $n=6$ qubits consisting of an $n_A = 3$-qubit memory subregister and of an  $n_B = 3$-qubit output subregister. The parameterized unitary $U(x, \theta)$ implements the cascade of an \emph{encoding unitary} $U_{in}(x)$, and of a problem-dependent \emph{evolution unitary} $U_H (\theta)$. Specifically, in order to encode each input sample $x$ we use the encoding unitary $U_{in} (x)=I^A \otimes R_{in}(x)^{\otimes 3}$, which applies an identity tranformation to the memory subregister and an input-dependent Pauli-$Y$ rotation \cite{takaki2021learning}
\begin{align}
    R_{in} (x) = R_y (\text{arccos}(x))=\exp(-i\text{arccos}(x)Y/2)
\end{align}
to each qubit of the output register.  As shown in Fig.~\ref{UH},  the evolution unitary consists of a layer of single qubit rotation gates, each parameterized by the angles $\theta$ as in \cite{takaki2021learning}, followed by Hamiltonian dynamics. Thereby, after the rotation layer, the unitary transformation $\exp(-iH \Delta t)$ is applied using the fixed Hamiltonian
\begin{align}\label{Hamiltonian_circ}
    H = \sum_{i=1}^n a_i X_i + \sum_{i=1}^n \sum_{j=1}^{i-1} J_{ij} Z_i Z_J 
\end{align} with $\Delta t=0.17$. 
The coefficients $a_i, J_{ij}$ in (\ref{Hamiltonian_circ}) are drawn randomly from a uniform distribution as $a_i, J_{ij} \in [-1,1]$, and they are fixed during training. In contrast, the rotation angles are optimized using the gradient-free optimizer COBLYA \cite{2020SciPy-NMeth}.  
We emphasize that the encoding circuit could be chosen differently, and that the architecture has not been optimized. For example, we may be able to improve the performance of the models by encoding orthogonal polynomials or using trainable encoding strategies \cite{lloyd2020quantum}. At the measurement part, we measure Z-expectation value of each qubit in the output register and take $z_t$ as their average multiplied by a real coefficient $c$ for the (TWI-)QRNN.

\subsubsection{Gating Mechanism via a Classical RNN}
For the classical RNN in the proposed models, we do not use hidden layers and we directly optimize the gating parameters $W = \{W_x, W_h, W^h_x, W^h_h\}$ determining the hyperparameter in \eqref{phi_hyper}  using gradient descent as discussed in Sec. III-D. The optimization of the circuit and gating parameters is carried over in an iterative manner.

\subsubsection{Classical Benchmark}
As a classical benchmark, we adopt a standard temporal model comprised of a LSTM RNN layer followed by a dense layer.  To ensure a meaningful comparison with quantum models,  we set the number of trainable parameters in the gating mechanism of the TWI-QRNN to match the number of parameters in forget gate of the LSTM, namely $4$ trainable parameters and bias terms, and the number of trainable parameters in the circuit of the TWI-QRNN to match the number of parameters in the input gate, namely $3$ trainable parameters. Note that, the LSTM has additional parameters in the output gate and the dense layer, for a total of $14$ parameters. This choice ensures that the number of parameters is the same in modules that appear in both models since the TWI-QRNN does not have an output gate/dense layer. The model is trained using gradient descent, using the Adam optimiser over $2000$ training epochs with a learning rate $0.001$. We use the mean squared error loss as the optimisation objective.

\subsection{Results}

\begin{figure*}[tbp]
\centering
\subfigure{\includegraphics[width=0.4\textwidth]{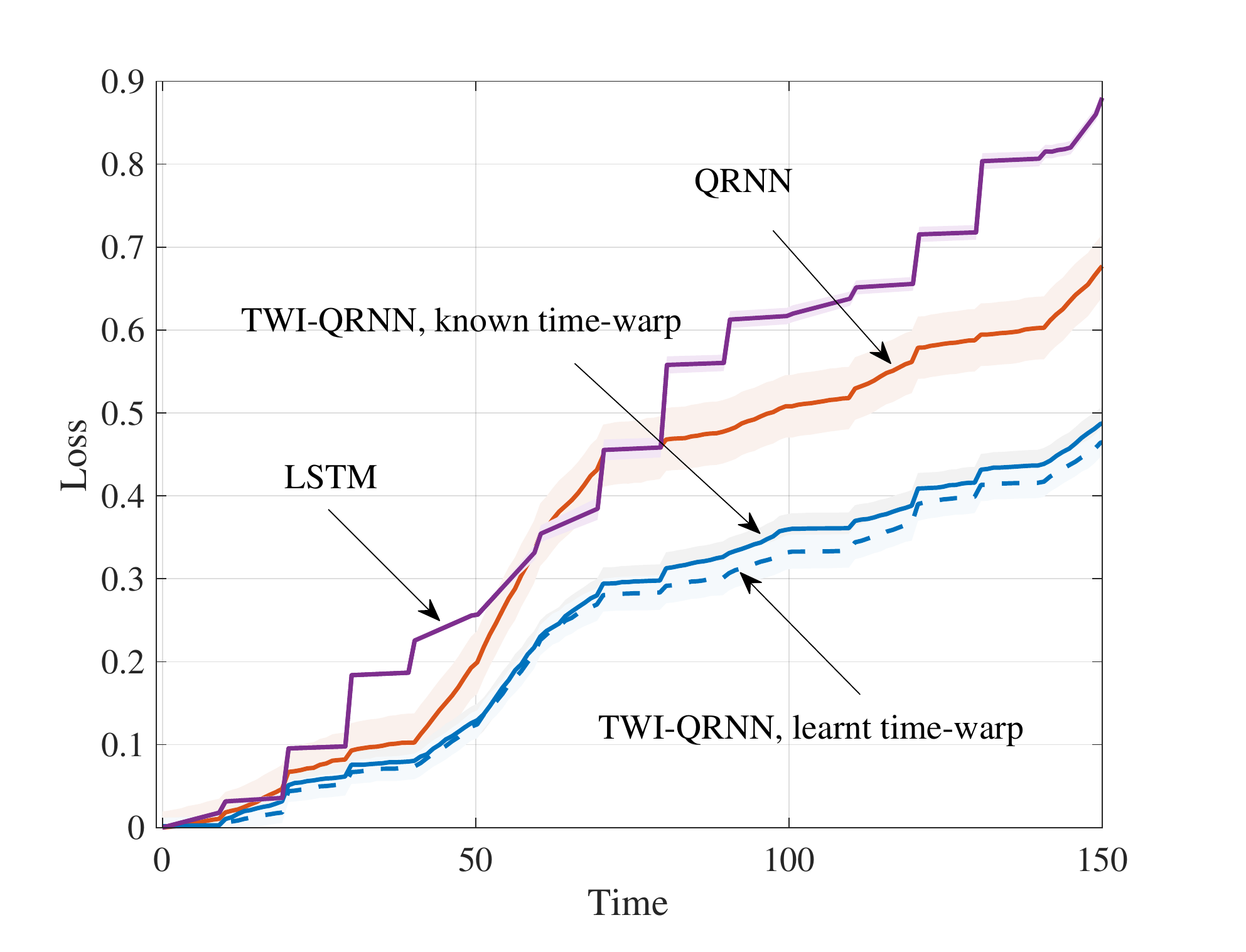}}
\subfigure{\includegraphics[width=0.39\textwidth]{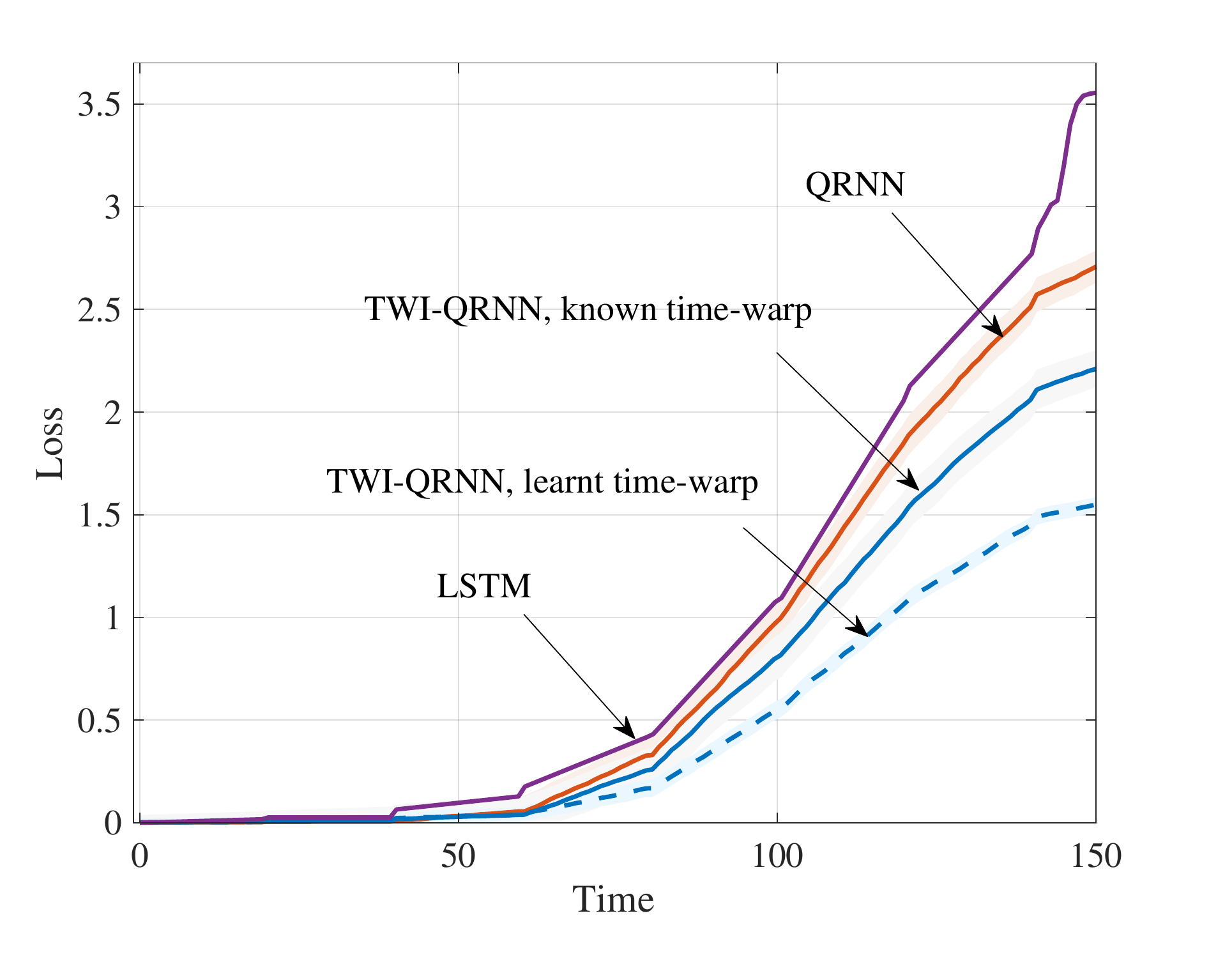}}
\caption{Cumulative quadratic loss for the task of remembering a time-warped cosine wave as a function of time for time-warping parameters $a = 0.1$ (left) and $a = 0.05$ (right).}
\label{res1}
\end{figure*}

\subsubsection{Remembering a Cosine Wave}

For the first task -- remembering the last sample of a, possibly time-warped, cosine wave -- we show the accumulated prediction loss over time in Fig.~\ref{res1}. The training algorithm is run for the first $50$ samples (not shown), and the error is accumulated over the subsequent $150$ samples. We use $5$ trials, and we average over the random initialization of the trainable parameters, as well as the Hamiltonian coefficients in \eqref{Hamiltonian_circ}.

\begin{figure*}[tbp]
\centering
\subfigure{\includegraphics[width=0.4\textwidth]{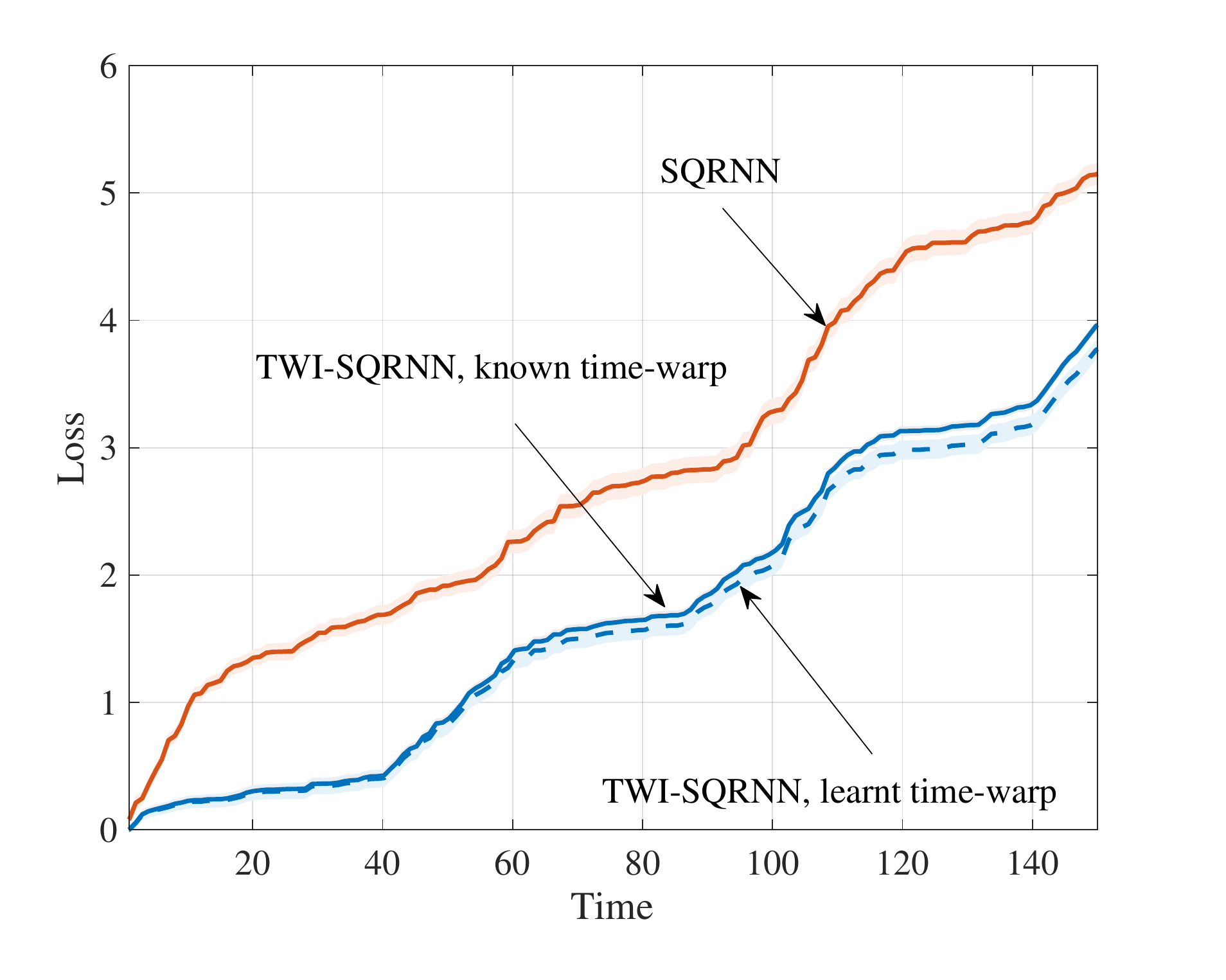}}
\subfigure{\includegraphics[width=0.39\textwidth]{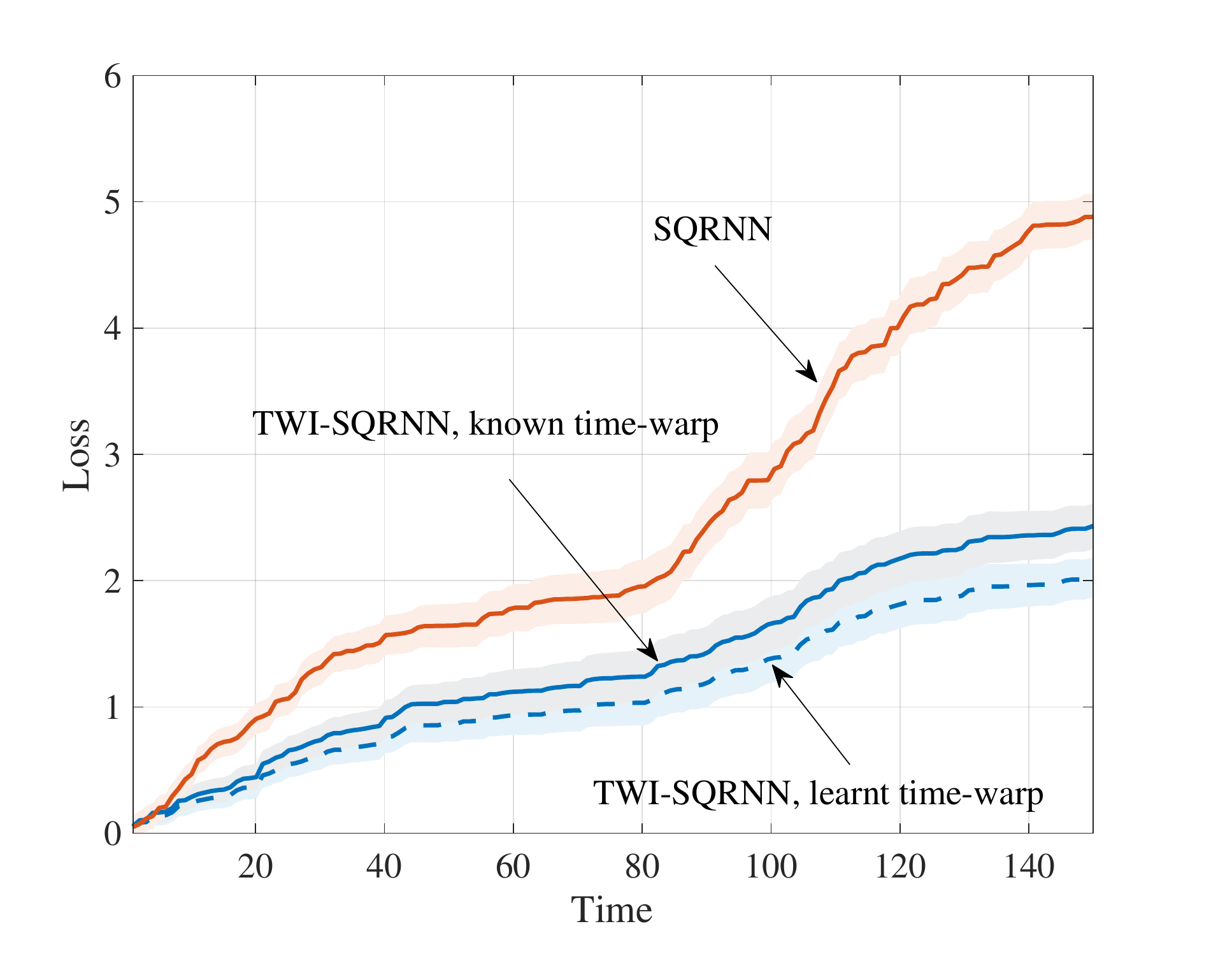}}
\caption{Cumulative quadratic loss for the task of remembering a time-warped cosine wave as a function of time for time-warping parameters $a = 0.1$ (left) and $a = 0.05$ (right). Unlike Fig. \ref{res1}, which considers deterministic predictors obtained via expected values of an observable, this figure consider stochastic, one-shot, predictors based on a single measurement output (see Sec. 4).}
\label{res_prob}
\end{figure*}

We consider the case when the warping parameter $a$ is known a priori, as well as the more challenging case in which parameter $a$ is not known and is learnt via the proposed RNN-based adaptive gating mechanism. For both considered value of $a$, namely $a=0.1$ and $a=0.05$, the proposed TWI-QRNN model is seen to consistently achieve a significantly lower loss as compared with the standard QRNN. This demonstrates the robustness of TWI-QRNNs to linear time warps. The gap in performance between TWI-QRNNs and QRNNs widens as time warping becomes more pronounced by having $a$ increase from $a = 0.1$ to $a = 0.05$. Similar conclusions can be drawn for the stochastic model, as shown in Fig.~\ref{res_prob}. In fact, TWI-SQRNN outperforms the SQRNN model for both values of $a$. In addition, the TWI-QRNN is seen to accrue a lower loss compared to the classical LSTM benchmark.

Overall, for both models, the gating mechanism is seen to be capable of adapting the value of $a$, achieving comparable loss to the case when $a$ is known a priori. In fact, the loss obtained with adaptive gating is lower than that with known parameter $a$. {\color{black}{This suggests that the adaptive gating mechanism, which is further investigated in Appendix A,  may compensate for errors in the approximations adopted in Sec. 3. }}

\subsubsection{Predicting Spin Dynamics} 

\begin{figure*}[tbp]
\centering
\includegraphics[width=0.5\textwidth]{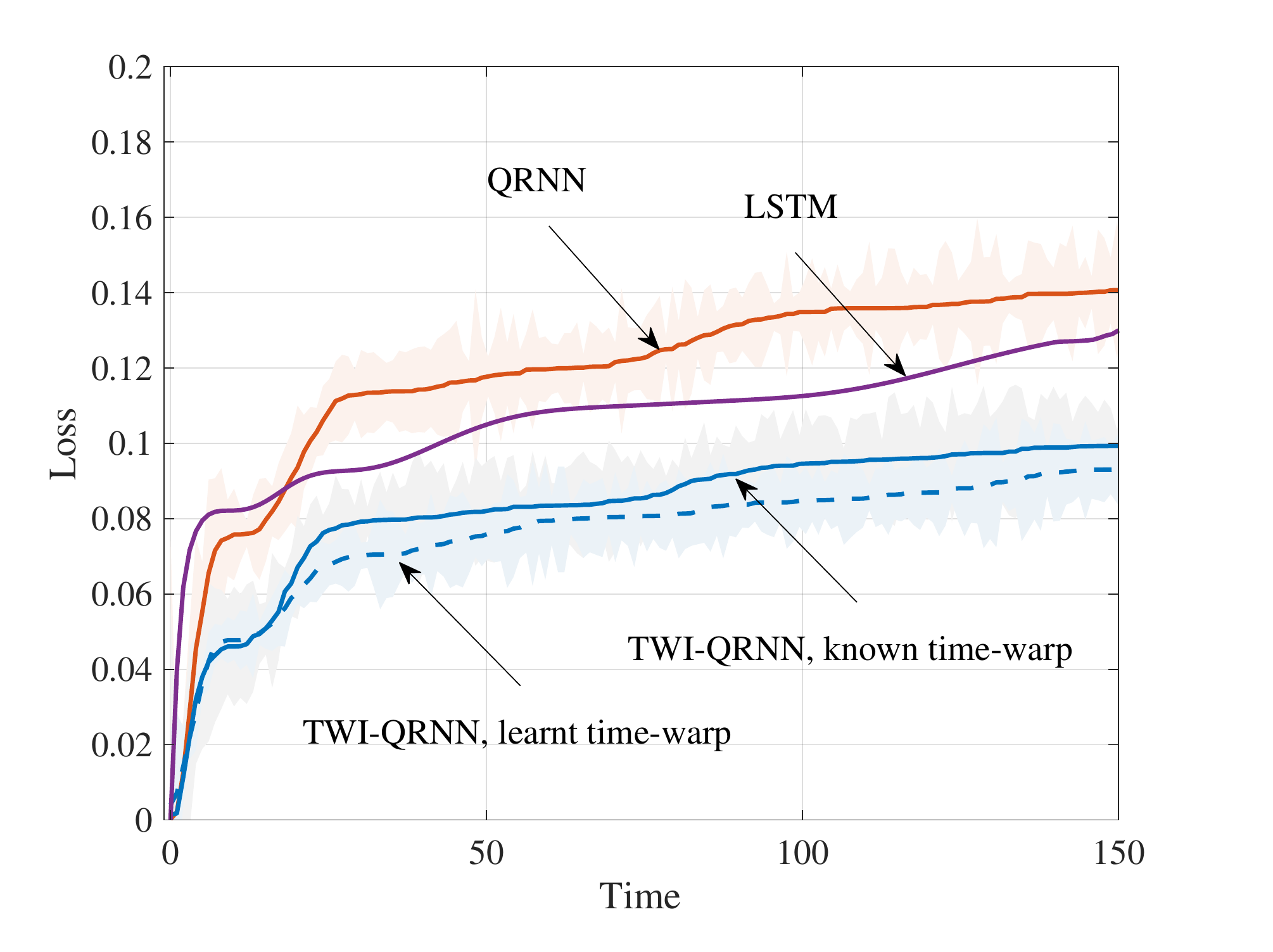}\vspace{-0.4cm}
\caption{Cumulative quadratic loss for the task of predicting time-warped spin dynamics as a function of time for time-warping function $c(t)=\sqrt(t)$.}
\vspace{-0.5cm}
\label{res3}
\end{figure*}

In a manner similar to the previous experiment, the accumulated quadratic loss for the problem of predicting spin dynamics over time is shown in Fig.~\ref{res3}. We consider again both the ideal case in which the time-warping function $c(t) = \sqrt{t}$  is known a priori, and the more practical case in which the time transformation is not known and is learnt via the RNN-based adaptive gating mechanism. 

The accrued loss accumulates more quickly in the earlier time steps as the warping is more prominent when the continuous time variable $t$ is smaller. However, the proposed model is  seen to be capable of resisting even non-linear warping transformations, accruing much lower prediction loss over time compared to the conventional QRNN model. This confirms that the gating mechanism can approximate the warping derivative and achieve low prediction loss, encouraging the use of the proposed model for prediction of phenomena characterized by quantum dynamics. In a manner similar to the previous experiment, the TWI-QRNN also performs better than the classical LSTM benchmark.

\section{Discussion and Future Work}

In this paper, we have studied quantum models that process temporal data. We have showed that postulating invariance to time transformations in the data,  hence taking invariance to time warping as an axiom, necessarily leads to an adaptive gate-like mechanism in quantum recurrent models. We derived a novel quantum model class from first principles, which was experimentally seen to be capable of resisting time warping transformations. We have also provided examples in which the proposed quantum model class outperforms a classical LSTM RNN benchmark with comparable capacity in the corresponding modules in terms of number of parameters. In this regard, we caution against deriving a general conclusion based on these results. Current state-of-the-art machine learning models fit large datasets with models with large capacity which is not plausible with current quantum devices due to issues associated with small numbers of qubits and data loading, as well as trainability challenges due to barren plateaus. However, when the machine learning practitioner is concerned with learning from limited temporal data, as the results of this work show, the proposed model is the better candidate to tackle the problem.

Various future research directions arise. For example, it is an open question how to design an efficient TWI-QRNN models with a fully quantum gating mechanism. This is particularly useful for running the models on quantum hardware, rather than simulators, and it is also left for future work. 
An interesting direction could be the investigation of the memory cost of time warping-invariant quantum recurrent models, since quantum models have been shown to describe temporal sequences with reduced memory \cite{yang2021provable,elliott2020extreme}. Given the performance of the proposed model on prediction tasks, we expect it to be advantageous for tasks related to quantum phenomena, with extensions possibly operating directly on quantum data \cite{huang2022quantum}.  

\section*{Acknowledgements}
We would like to thank Johannes Bausch and Danilo Rezende for reviewing the paper prior to submission. OS acknowledges funding from the European Research Council (ERC) under the European Union’s Horizon 2020 Research and Innovation Program (Grant Agreement No. 725731) and from the EPSRC (EP/W024101/1). LB acknowledges funding from the
U.S. Department of Energy, Office of Science, National
Quantum Information Science Research Centers, Superconducting Quantum Materials and Systems Center
(Contract No. DE-AC02-07CH11359).

\appendix

{\color{black}{\section{Further Experiments}

\begin{figure*}[tbp]
\centering
\subfigure{\includegraphics[width=0.4\textwidth]{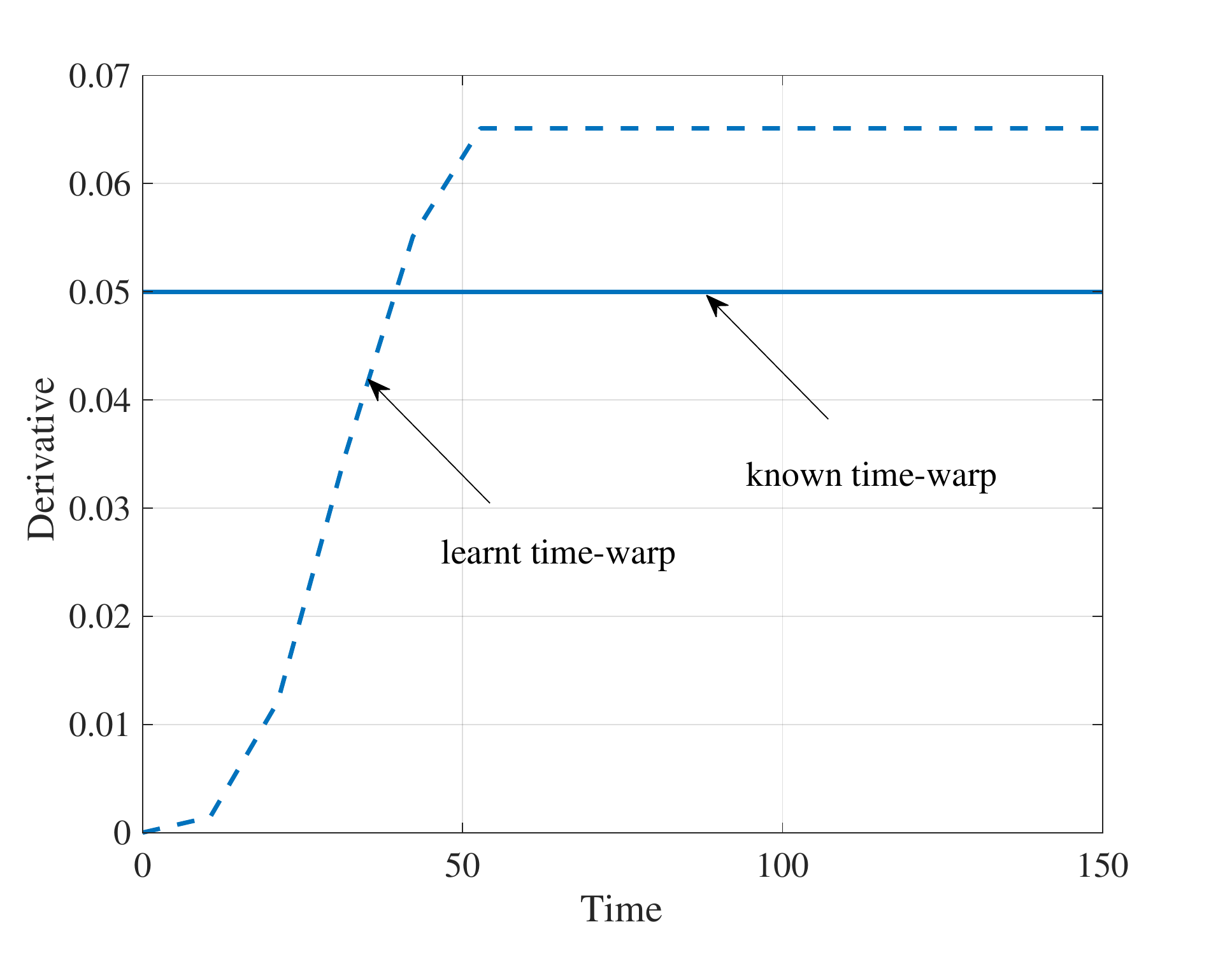}}
\subfigure{\includegraphics[width=0.4\textwidth]{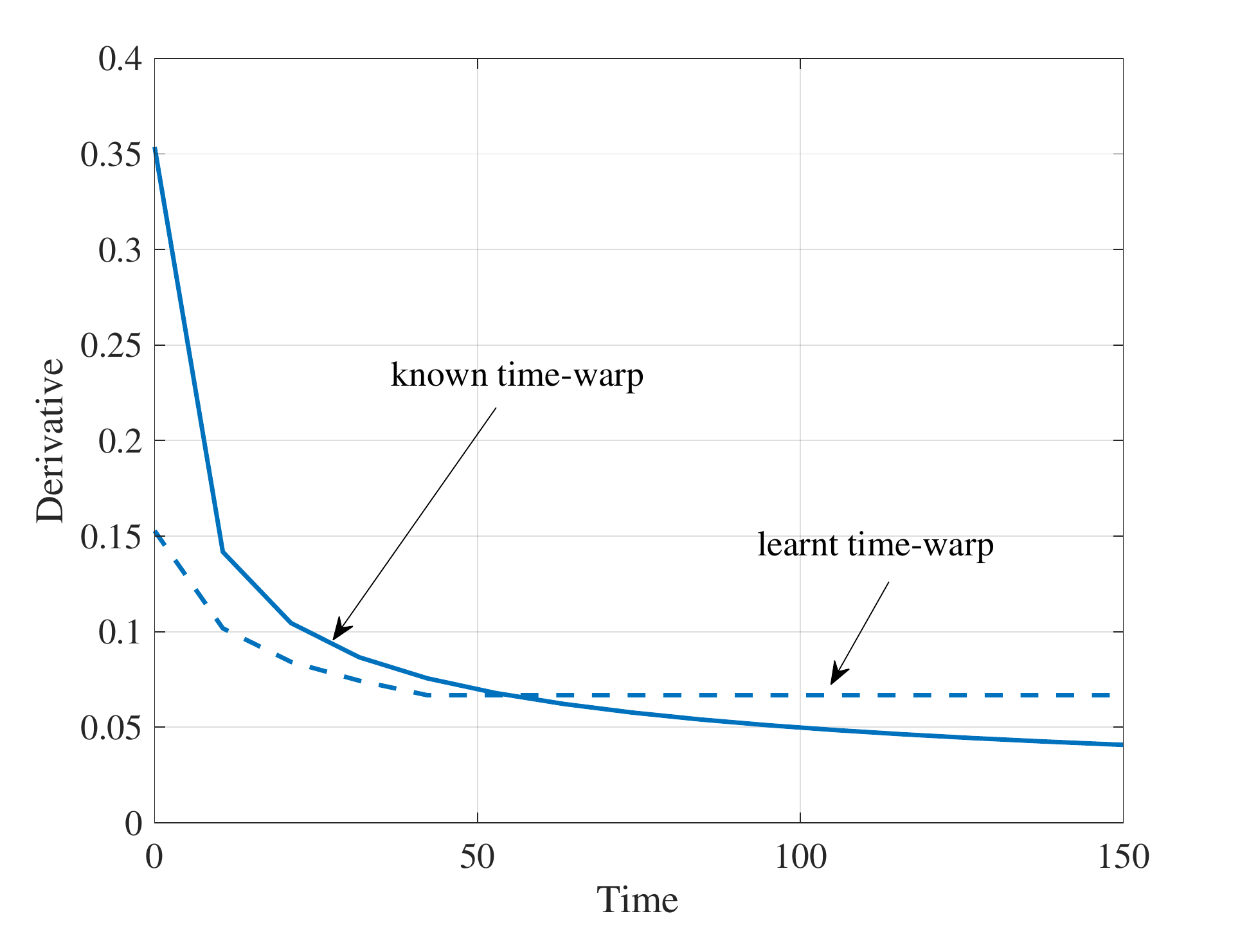}}
\caption{(left) Known time-warping derivative $dc(t)/dt$ and learnt derivative $\alpha_t$ in (\ref{derivative}) for linear warping $c(t) = a t$; and (right) for non-linear warping $c(t) = \sqrt{t}$.}
\label{fig:alpha_experiments}
\end{figure*}

In this appendix, we provide further insights on the extent to which the trained RNN fits the time-warping derivative $d c(t) / dt$. As demonstrated in Sec. 5 via experiments, the choice $\alpha_t \approx d c(t) / dt$, which is motivated by the heuristic derivation of the TWI-QRNN model in Sec. 3.2, is generally suboptimal in terms of performance loss. In fact, as seen in Fig. 8 and Fig. 9, the performance obtained with the ``known'' time-warp $\alpha_t \approx d c(t) / dt$ is improved by the `learnt'' time-warp  $\alpha_t$ in (\ref{derivative}) obtained via the RNN.

To this end, Fig.~\ref{fig:alpha_experiments} illustrates the known time-warping derivative $d c(t) / dt$ and learnt time-warping derivatives $\alpha_t$ in (\ref{derivative}) for the two examples studied in Sec. 5, which involve linear and non-linear warping. Whilst the output $\alpha_t$ of the RNN is seen to approximate the derivative $d c(t) / dt$, in both cases the match is not perfect. Together with the observations made in Fig. 8 and Fig. 9, this confirms that setting $\alpha = d c(t) / dt$ is generally not optimal, and the RNN may effectively compensate for the some of the approximation errors introduced in deriving the model in \eqref{eq:scaleinvariance}.}}


\section{Alternative Derivation of TWI-QRNNs}

In this section, we present an alternative derivation of the TWI-QRNN model that follows directly the steps in \cite{tallec2018can}. To start, we introduce a continuous-time version of the update (\ref{eq:cq1}) via the approximation
\begin{align}\label{TaylorQRNN}
    \rho(t+\delta t) & \approx \rho(t) + \frac{\textrm{d} \rho(t)}{\textrm{d} t} \delta t,
\end{align} where we have used $\rho(t)$ to denote a continuous-time version of the density matrix and $\delta t >0$ is a time interval. The update (\ref{eq:cq1}) can be viewed as the standard \emph{Euler approximation} of (\ref{TaylorQRNN}). Accordingly, under suitable assumptions, the accuracy of the approximation is proportional to $\delta t$ (see, e.g., \cite[Chapter 4]{meyn2022control}).  It is emphasized that (\ref{TaylorQRNN}) is not meant to be a description of the continuous-time dynamics of the system, which is dictated by the Liouville–von Neumann equation \cite{peres1997quantum}, but only a stepping stone in the derivation of time warping-invariant discrete-time models.

From (\ref{eq:cq1}), setting $\delta t=1$, we then have \begin{align}\label{TaylorQRNNone}
    \frac{\textrm{d} \rho(t)}{\textrm{d} t} = V(x(t), \theta) \rho(t) V(x(t), \theta)^{\dagger} - \rho(t),
\end{align}where we have kept the dependence of the unitary $V(x(t), \theta)$ on time $t$ implicit. Let us now apply a time-warping operation so that the input signal observed by the QRNN at time $t$ is given by $x(c(t))$. Using \eqref{TaylorQRNNone}, the output density satisfies the equality
\begin{align}
    \frac{\textrm{d} \rho(c(t))}{\textrm{d} c(t)} &= V(x(c(t)), \theta) \rho(c(t)) V(x(c(t)), \theta)^{\dagger} \nonumber\\
    &- \rho(c(t)).
\end{align}
Using the chain rule of differentiation, this in turn yields
\begin{align}\label{derivatQRNN}
    \frac{\textrm{d} \rho(c(t))}{\textrm{d} t} &= \frac{\textrm{d} \rho(c(t))}{\textrm{d} c(t)} \cdot \frac{\textrm{d} c(t)}{\textrm{d} t} \nonumber\\
    &=  \frac{\textrm{d} c(t)}{\textrm{d} t}[V(x(c(t)), \theta) \rho(c(t)) V(x(c(t)), \theta)^{\dagger} \nonumber\\
    &- \rho(c(t))].
\end{align}
To obtain a relationship between density matrices $\rho(c(t+1))$ and $\rho(c(t))$, we take a Taylor expansion on the time-warped model, whereby we obtain
\begin{align}\label{TaylorQRNNwa}
    \rho(c(t+\delta t)) & \approx \rho(c(t)) + \frac{\textrm{d} \rho(c(t))}{\textrm{d} t} \delta t.
\end{align}
By plugging in \eqref{derivatQRNN}, and taking $\delta t = 1$, this yields
\begin{align}\label{TwIQRNNcont}
    &\rho(c(t+1)) = \left(1-\frac{\textrm{d} c(t)}{\textrm{d} t}\right)\rho(c(t)) \nonumber\\
    &+ \frac{\textrm{d} c(t)}{\textrm{d} t}  V(x(c(t)), \theta) \rho(c(t)) V(x(c(t)), \theta)^{\dagger}.
\end{align}

Following the arguments in \cite{tallec2018can} and \cite{bronstein2021geometric}, we finally return to a discrete-time model as follows. From the point of view of a discrete QRNN model under time warping, the time $c(t)$ corresponds to discrete time $t$, the value $x(c(t))$ to $x_t$, and $\rho(c(t))$ to $\rho_t$. Therefore, a discrete-time version of \eqref{TwIQRNNcont} can be defined as (\ref{lQRNN}).

\bibliographystyle{IEEEtran}
\bibliography{litdab.bib}

\begin{thebibliography}{10}
\providecommand{\url}[1]{#1}
\csname url@samestyle\endcsname
\providecommand{\newblock}{\relax}
\providecommand{\bibinfo}[2]{#2}
\providecommand{\BIBentrySTDinterwordspacing}{\spaceskip=0pt\relax}
\providecommand{\BIBentryALTinterwordstretchfactor}{4}
\providecommand{\BIBentryALTinterwordspacing}{\spaceskip=\fontdimen2\font plus
\BIBentryALTinterwordstretchfactor\fontdimen3\font minus
  \fontdimen4\font\relax}
\providecommand{\BIBforeignlanguage}[2]{{%
\expandafter\ifx\csname l@#1\endcsname\relax
\typeout{** WARNING: IEEEtran.bst: No hyphenation pattern has been}%
\typeout{** loaded for the language `#1'. Using the pattern for}%
\typeout{** the default language instead.}%
\else
\language=\csname l@#1\endcsname
\fi
#2}}
\providecommand{\BIBdecl}{\relax}
\BIBdecl

\bibitem{chowdhary2020natural}
K.~Chowdhary, ``Natural language processing,'' \emph{Fundamentals of artificial
  intelligence}, pp. 603--649, 2020.

\bibitem{hirschberg2015advances}
J.~Hirschberg and C.~D. Manning, ``Advances in natural language processing,''
  \emph{Science}, vol. 349, no. 6245, pp. 261--266, 2015.

\bibitem{georgescu2014quantum}
I.~M. Georgescu, S.~Ashhab, and F.~Nori, ``Quantum simulation,'' \emph{Reviews
  of Modern Physics}, vol.~86, no.~1, p. 153, 2014.

\bibitem{mitarai2018quantum}
K.~Mitarai, M.~Negoro, M.~Kitagawa, and K.~Fujii, ``Quantum circuit learning,''
  \emph{Physical Review A}, vol.~98, no.~3, p. 032309, 2018.

\bibitem{graves2012long}
A.~Graves, ``Long short-term memory,'' \emph{Supervised sequence labelling with
  recurrent neural networks}, pp. 37--45, 2012.

\bibitem{graves2013speech}
A.~Graves, A.-r. Mohamed, and G.~Hinton, ``Speech recognition with deep
  recurrent neural networks,'' in \emph{2013 IEEE international conference on
  acoustics, speech and signal processing}.\hskip 1em plus 0.5em minus
  0.4em\relax Ieee, 2013, pp. 6645--6649.

\bibitem{tallec2018can}
C.~Tallec and Y.~Ollivier, ``Can recurrent neural networks warp time?''
  \emph{arXiv preprint arXiv:1804.11188}, 2018.

\bibitem{bronstein2021geometric}
M.~M. Bronstein, J.~Bruna, T.~Cohen, and P.~Veli{\v{c}}kovi{\'c}, ``Geometric
  deep learning: Grids, groups, graphs, geodesics, and gauges,'' \emph{arXiv
  preprint arXiv:2104.13478}, 2021.

\bibitem{velivckovic2017graph}
P.~Veli{\v{c}}kovi{\'c}, G.~Cucurull, A.~Casanova, A.~Romero, P.~Lio, and
  Y.~Bengio, ``Graph attention networks,'' \emph{arXiv preprint
  arXiv:1710.10903}, 2017.

\bibitem{krizhevsky2012imagenet}
A.~Krizhevsky, I.~Sutskever, and G.~E. Hinton, ``Imagenet classification with
  deep convolutional neural networks,'' \emph{Advances in neural information
  processing systems}, vol.~25, pp. 1097--1105, December, 2012.

\bibitem{larocca2022group}
M.~Larocca, F.~Sauvage, F.~M. Sbahi, G.~Verdon, P.~J. Coles, and M.~Cerezo,
  ``Group-invariant quantum machine learning,'' \emph{arXiv preprint
  arXiv:2205.02261}, 2022.

\bibitem{meyer2022exploiting}
J.~J. Meyer, M.~Mularski, E.~Gil-Fuster, A.~A. Mele, F.~Arzani, A.~Wilms, and
  J.~Eisert, ``Exploiting symmetry in variational quantum machine learning,''
  \emph{arXiv preprint arXiv:2205.06217}, 2022.

\bibitem{ragone2022representation}
M.~Ragone, P.~Braccia, Q.~T. Nguyen, L.~Schatzki, P.~J. Coles, F.~Sauvage,
  M.~Larocca, and M.~Cerezo, ``Representation theory for geometric quantum
  machine learning,'' \emph{arXiv preprint arXiv:2210.07980}, 2022.

\bibitem{verdon2019learning}
G.~Verdon, M.~Broughton, J.~R. McClean, K.~J. Sung, R.~Babbush, Z.~Jiang,
  H.~Neven, and M.~Mohseni, ``Learning to learn with quantum neural networks
  via classical neural networks,'' \emph{arXiv preprint arXiv:1907.05415},
  2019.

\bibitem{verdon2019quantum}
G.~Verdon, T.~McCourt, E.~Luzhnica, V.~Singh, S.~Leichenauer, and J.~Hidary,
  ``Quantum graph neural networks,'' \emph{arXiv preprint arXiv:1909.12264},
  2019.

\bibitem{ai2022decompositional}
X.~Ai, Z.~Zhang, L.~Sun, J.~Yan, and E.~Hancock, ``Decompositional quantum
  graph neural network,'' \emph{arXiv preprint arXiv:2201.05158}, 2022.

\bibitem{mernyei2022equivariant}
P.~Mernyei, K.~Meichanetzidis, and I.~I. Ceylan, ``Equivariant quantum graph
  circuits,'' in \emph{International Conference on Machine Learning}.\hskip 1em
  plus 0.5em minus 0.4em\relax PMLR, 2022, pp. 15\,401--15\,420.

\bibitem{nguyen2022theory}
Q.~T. Nguyen, L.~Schatzki, P.~Braccia, M.~Ragone, P.~J. Coles, F.~Sauvage,
  M.~Larocca, and M.~Cerezo, ``Theory for equivariant quantum neural
  networks,'' \emph{arXiv preprint arXiv:2210.08566}, 2022.

\bibitem{schatzki2022theoretical}
L.~Schatzki, M.~Larocca, F.~Sauvage, and M.~Cerezo, ``Theoretical guarantees
  for permutation-equivariant quantum neural networks,'' \emph{arXiv preprint
  arXiv:2210.09974}, 2022.

\bibitem{pesah2021absence}
A.~Pesah, M.~Cerezo, S.~Wang, T.~Volkoff, A.~T. Sornborger, and P.~J. Coles,
  ``Absence of barren plateaus in quantum convolutional neural networks,''
  \emph{Physical Review X}, vol.~11, no.~4, p. 041011, 2021.

\bibitem{bausch2020recurrent}
J.~Bausch, ``Recurrent quantum neural networks,'' \emph{Advances in neural
  information processing systems}, vol.~33, pp. 1368--1379, 2020.

\bibitem{takaki2021learning}
Y.~Takaki, K.~Mitarai, M.~Negoro, K.~Fujii, and M.~Kitagawa, ``Learning
  temporal data with a variational quantum recurrent neural network,''
  \emph{Physical Review A}, vol. 103, no.~5, p. 052414, 2021.

\bibitem{elliott2022quantum}
T.~J. Elliott, M.~Gu, A.~J. Garner, and J.~Thompson, ``Quantum adaptive agents
  with efficient long-term memories,'' \emph{Physical Review X}, vol.~12,
  no.~1, p. 011007, 2022.

\bibitem{elliott2020extreme}
T.~J. Elliott, C.~Yang, F.~C. Binder, A.~J. Garner, J.~Thompson, and M.~Gu,
  ``Extreme dimensionality reduction with quantum modeling,'' \emph{Physical
  Review Letters}, vol. 125, no.~26, p. 260501, 2020.

\bibitem{yang2021provable}
C.~Yang, A.~Garner, F.~Liu, N.~Tischler, J.~Thompson, M.-H. Yung, M.~Gu, and
  O.~Dahlsten, ``Provable superior accuracy in machine learned quantum
  models,'' \emph{arXiv preprint arXiv:2105.14434}, 2021.

\bibitem{chen2020temporal}
J.~Chen, H.~I. Nurdin, and N.~Yamamoto, ``Temporal information processing on
  noisy quantum computers,'' \emph{Physical Review Applied}, vol.~14, no.~2, p.
  024065, 2020.

\bibitem{wellsfargo}
S.~Y.-C. Chen, D.~Fry, A.~Deshmukh, V.~Rastunkov, and C.~Stefanski, ``Reservoir
  computing via quantum recurrent neural networks,'' 2022.

\bibitem{banchi2018modelling}
L.~Banchi, E.~Grant, A.~Rocchetto, and S.~Severini, ``Modelling non-markovian
  quantum processes with recurrent neural networks,'' \emph{New Journal of
  Physics}, vol.~20, no.~12, p. 123030, 2018.

\bibitem{sharma2022trainability}
K.~Sharma, M.~Cerezo, L.~Cincio, and P.~J. Coles, ``Trainability of dissipative
  perceptron-based quantum neural networks,'' \emph{Physical Review Letters},
  vol. 128, no.~18, p. 180505, 2022.

\bibitem{Heimann2022LearningCO}
D.~Heimann, G.~Schonhoff, and F.~Kirchner, ``Learning capability of
  parametrized quantum circuits,'' 2022.

\bibitem{cao2017quantum}
Y.~Cao, G.~G. Guerreschi, and A.~Aspuru-Guzik, ``Quantum neuron: an elementary
  building block for machine learning on quantum computers,'' \emph{arXiv
  preprint arXiv:1711.11240}, 2017.

\bibitem{schuld2021machine}
M.~Schuld and F.~Petruccione, \emph{Machine learning with quantum
  computers}.\hskip 1em plus 0.5em minus 0.4em\relax Springer, 2021.

\bibitem{banchi2021measuring}
L.~Banchi and G.~E. Crooks, ``Measuring analytic gradients of general quantum
  evolution with the stochastic parameter shift rule,'' \emph{Quantum}, vol.~5,
  p. 386, 2021.

\bibitem{ciccarello2022quantum}
F.~Ciccarello, S.~Lorenzo, V.~Giovannetti, and G.~M. Palma, ``Quantum collision
  models: Open system dynamics from repeated interactions,'' \emph{Physics
  Reports}, vol. 954, pp. 1--70, 2022.

\bibitem{williams1992simple}
R.~J. Williams, ``Simple statistical gradient-following algorithms for
  connectionist reinforcement learning,'' \emph{Machine learning}, vol.~8,
  no.~3, pp. 229--256, 1992.

\bibitem{solomon2017pseudo}
G.~Solomon and L.~Weissfeld, ``Pseudo maximum likelihood approach for the
  analysis of multivariate left-censored longitudinal data,'' \emph{Statistics
  in medicine}, vol.~36, no.~1, pp. 81--91, 2017.

\bibitem{2020SciPy-NMeth}
P.~Virtanen, R.~Gommers, T.~E. Oliphant, M.~Haberland, T.~Reddy, D.~Cournapeau,
  E.~Burovski, P.~Peterson, W.~Weckesser, J.~Bright, S.~J. {van der Walt},
  M.~Brett, J.~Wilson, K.~J. Millman, N.~Mayorov, A.~R.~J. Nelson, E.~Jones,
  R.~Kern, E.~Larson, C.~J. Carey, {\.I}.~Polat, Y.~Feng, E.~W. Moore,
  J.~{VanderPlas}, D.~Laxalde, J.~Perktold, R.~Cimrman, I.~Henriksen, E.~A.
  Quintero, C.~R. Harris, A.~M. Archibald, A.~H. Ribeiro, F.~Pedregosa, P.~{van
  Mulbregt}, and {SciPy 1.0 Contributors}, ``{{SciPy} 1.0: Fundamental
  Algorithms for Scientific Computing in Python},'' \emph{Nature Methods},
  vol.~17, pp. 261--272, 2020.

\bibitem{lloyd2020quantum}
S.~Lloyd, M.~Schuld, A.~Ijaz, J.~Izaac, and N.~Killoran, ``Quantum embeddings
  for machine learning,'' \emph{arXiv preprint arXiv:2001.03622}, 2020.

\bibitem{huang2022quantum}
H.-Y. Huang, M.~Broughton, J.~Cotler, S.~Chen, J.~Li, M.~Mohseni, H.~Neven,
  R.~Babbush, R.~Kueng, J.~Preskill \emph{et~al.}, ``Quantum advantage in
  learning from experiments,'' \emph{Science}, vol. 376, no. 6598, pp.
  1182--1186, 2022.

\bibitem{meyn2022control}
S.~Meyn, \emph{Control Systems and Reinforcement Learning}.\hskip 1em plus
  0.5em minus 0.4em\relax Cambridge University Press, 2022.

\bibitem{peres1997quantum}
A.~Peres, \emph{Quantum theory: concepts and methods}.\hskip 1em plus 0.5em
  minus 0.4em\relax Springer, 1997.

\end{thebibliography}

\end{document}